\documentclass{iopart} 

\usepackage{iopams} 
\usepackage{graphicx}
\usepackage{epsfig,epstopdf} 
\usepackage{hyperref} 

\renewcommand{\Re}{{\rm Re}}
\renewcommand{\Im}{{\rm Im}}
\newcommand{\mathwith}{\quad\mbox{with}\quad}
\newcommand{\mathand}{\quad\mbox{and}\quad}
\newcommand{\av}[1]{\langle{#1}\rangle}

\usepackage[square,sort&compress]{natbib}

\begin{document}
\title{Quantum Simulations of Extended Hubbard Models with Dipolar Crystals}
\author{M~Ortner$^{1,2}$, A~Micheli$^{1,2}$, G~Pupillo$^{1,2}$, P~Zoller$^{1,2}$}
\address{$^1$Institute for Theoretical Physics, University of Innsbruck, A-6020 Innsbruck, Technikerstrasse 25, Austria}
\address{$^2$Institute for Quantum Optics and Quantum Information of the
Austrian Academy of Sciences, A-6020 Innsbruck, Austria}
\eads{michael.ortner@uibk.ac.at}

\date{\today}

\begin{abstract}
In this paper we study the realization of lattice models in mixtures of atomic and dipolar molecular quantum gases. We consider a situation where polar molecules form a self-assembled dipolar lattice, in which atoms or molecules of a second species can move and scatter. We describe the system dynamics in a master equation approach in the Brownian motion limit of slow particles and fast phonons, which we find appropriate for our system. In a wide regime of parameters, the reduced dynamics of the particles leads to physical realizations of extended Hubbard models with tuneable long-range interactions mediated by crystal phonons. This extends the notion of quantum simulation of strongly correlated systems with cold atoms and molecules to include phonon-dynamics, where all coupling parameters can be controlled by external fields.
\end{abstract}

\maketitle


\newpage
\section{Introduction}\label{section:1}

The recent preparation of cold ensembles of homonuclear and heteronuclear molecules in the electronic and vibrational ground state \cite{BookPolarMol09,
Stwalley04, Stwalley04FB, DeMille04, DeMille05, Weidemueller06, Ye08Science, Jin08NPhys, Weidemuller08PRL100, Weidemuller08PRL101, Weidemuller08JCP, Gould08SCIENCE,
Denschlag07PRL, Grimm08, Denschlag08NPHYS, Denschlag08PRL, Pillet01, Pillet02,
Doyle98, Pillet98, Meijer00, Meijer01, Hinds04, Ye04, Rempe04, Rempe05, Meijer05, Ye07, Ye07PRA,
Doyle07}  has opened the door to a new chapter in the theoretical and experimental study of trapped quantum degenerate gases \cite{Bohn03,Krems05,Krems06, Bohn05, DeMille02QC, Micheli06NPHYS,Buechler07NPHYS, Pillet06, Rempe06NPHYS, Stwalley07, Meijer07PRL, Rempe08SCIENCE, Jin03Nature, Jin03Nature2, Jin08PRA, DeMille08, Grimm07, Ye08PRLMR, Yelin06, Ticknor09Arxiv, Carr08Arxiv, BuechlerPRL2007, GorshkovPRL08}. Heteronuclear  molecules, in particular, have large electric dipole moments associated with rotational excitations, and the new aspect in quantum gases of cold polar molecules is thus the large, anisotropic dipole-dipole interactions between the molecules which can be manipulated via external DC and AC fields in the microwave regime~\cite{Bohn03, Krems05, Krems06, Bohn05, BuechlerPRL2007, GorshkovPRL08, MicheliPRA2007}.
In combination with reduced trapping geometries, this promises the realization of novel quantum phases and quantum phase transitions, for example in the case of bosons the transition from a dipolar superfluid to a crystalline regime~\cite{BuechlerPRL2007}.  The theory of dipolar quantum gases, and various aspects of strongly correlated systems of polar molecules has been recently reviewed by Baranov~\cite{BaranovPhysRep08} and by Pupillo et al.~\cite{PupilloReview2008}

Below we will  extend these theoretical studies to {\em mixtures of atomic and dipolar molecular quantum gases}. More specifically, we will be interested in a situation where polar molecules form a self-assembled dipolar lattice, in which atoms can move and scatter. This scenario leads to a new physical realization of a Hubbard model where atoms see the periodic structure provided by the crystal formed by the polar molecules. In contrast to the familiar case of the realization of Hubbard models with cold atoms in optical lattices, where standing laser light waves produce a {\em fixed} periodic external lattice potential, self-assembled dipolar lattices have their own lattice dynamics represented by phonons.  Thus atoms moving in dipolar crystals give rise to Hubbard models which include both (i) atom-phonon couplings, and (ii) atom-atom interactions. This extends the notion of quantum simulation of strongly correlated systems with cold atoms to include phonon-dynamics, where both the atomic and phonon coupling parameters can be controlled by external fields. We note that atom - molecule mixtures arise naturally in photoassociation experiments, where a two-species atomic quantum degenerate gas is {\em partially} transferred to ground state molecules via formation of highlying Feshbach molecules, followed by a Raman transfer to the ground state. Similar Hubbard models result also in mixtures of two unbalanced species of polar molecules, where the first molecular species forms a crystal while the second species provides the extra particles hopping in the self-assembled dipolar lattice.

In a recent work~\cite{PupilloPRL2008}, we have shown that the dynamics of atoms and molecules embedded in dipolar crystals is conveniently described by a {\it polaronic} picture where the
particles are dressed by the crystal phonons. 
Standard treatments of polaron dynamics show a competition between coherent and incoherent hopping of a particle in the lattice. The former corresponds to tunneling of a particle from one site to to the next, while carrying the lattice distortion, without changing the phonon occupation. The latter corresponds to {\it thermally activated} particle hopping, related to incoherent hopping events where the number of phonons changes in the hop. That is, the polaron loses its phase coherence via the emission or absorption of phonons~\cite{Mahan}. The physics of polarons has a long history, dating back to the seminal work of Landau~\cite{Landau33}. Excellent reviews on the subject can be found e.g. in~\cite{Devreese1, Alexandrov, Kuper, Alexandrov2, Calvani, Fermi, Kleinert, Devreese2}. Here we take the simple approach of placing these coherent and incoherent processes in the natural framework of a master equation treatment of the system dynamics, where the phonons are treated as a thermal heat bath. We extend our work~\cite{PupilloPRL2008} by re-deriving the results of~\cite{Mahan} in the master equation context, and calculating additional corrections to the coherent and incoherent time evolution for our atomic and molecular mixtures. Since we find that in the latter the polaron dynamics is typically slow compared to the characteristic time evolution of the bath, we specialize to the Brownian motion limit of the master equation~\cite{Breuer,Carmichael}. We show that for the models of interest and low-enough temperatures, corrections to the coherent time evolution of the polaron system are small, and thus the dynamics of the dressed particles is well described by an effective extended Hubbard model in a wide range of realistic parameters.\\

The paper is organized as follows. In Section~\ref{section:2} below we derive the Brownian motion master equation for our system, providing explicit expressions for the coherent and incoherent polaron dynamics (details of the derivation can be found in \ref{AppC}). In Section~\ref{settingup} we review the realization of dipolar crystals in two and one dimensions. In Section~\ref{rdc} we provide details of the implementation of polaronic extended Hubbard models for two one-dimensional configurations with atoms and molecules.

\begin{figure}[t!]
\begin{flushright}
\includegraphics[width=.85\columnwidth]{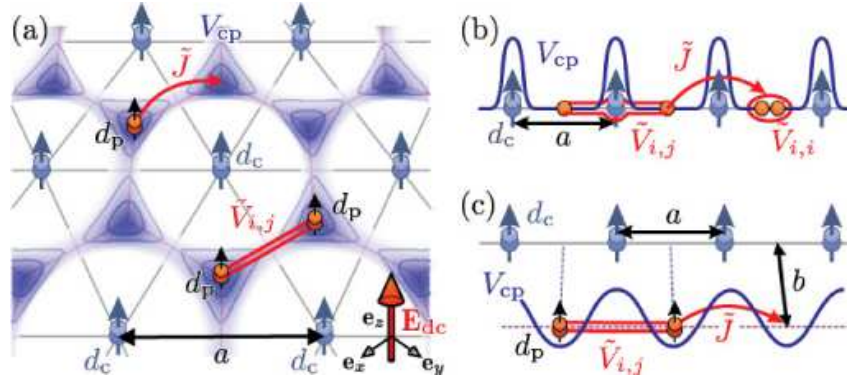}
\end{flushright}
\caption{\label{figs:fig1}A dipolar crystal of polar molecules in 2D
(a) and 1D (b,c) provides a periodic lattice $V_{{\rm cp}}$ for
extra atoms or molecules giving rise to a lattice model with hopping
$\tilde{J}$ and long-range interactions $\tilde{V}_{i,j}$ (see
text). (a) In 2D a triangular lattice is formed by polar molecules
with dipole moment $d_{{\rm c}}$ perpendicular to the plane. A
second molecular species with dipole moment $d_{{\rm p}}\ll d_{{\rm
c}}$ moves in the honeycomb lattice $V_{{\rm cp}}$ (darker shading
corresponds to deeper potentials). (b) A 1D setup with atoms scattering form a dipolar crystal with lattice spacing $a$. (c) A 1D dipolar crystal provides a periodic potential for a second molecular species moving in a parallel tube at distance $b$.}
\end{figure}

\section{Dynamics of particles trapped in a crystal of polar molecules}\label{section:2}
We show below in \Sref{rdc} and \ref{AppA} that in a wide range of system parameters the dynamics of atoms or molecules which move in a self-assembled lattice of dipoles is well described by the following Hamiltonian
\begin{eqnarray}
 H &=& -J\sum_{\langle i,j\rangle}c_i^{\dag}c_j  +\frac{1}{2}\sum_{i,j}V_{i,j}c_i^{\dag}c_j^{\dag}c_jc_i \nonumber\\
&&+ \sum_{{\bf q},\lambda}\hbar\omega_{{\bf q},\lambda}a_{{\bf q},\lambda}^{\dag}a_{{\bf q},\lambda}+\sum_{{\bf q},\lambda,j}M_{{\bi q},\lambda}e^{i{\bf q}{\bf r}_j^0}c_j^{\dag}c_j(a_{{\bf q},\lambda}+a_{-{\bf q},\lambda}^{\dag}),\label{eq:1}
\end{eqnarray}
where the brackets $\langle \rangle$ indicate sums over nearest neighbors. The latter is a single-band Hubbard model for the particles coupled to the acoustic phonons of the lattice. The first two terms on the r.h.s. of \eref{eq:1} describe the hopping of particles between neighboring sites of the lattice with a tunneling rate $J$, and the density-density interactions with strength $V_{i,j}$ for particles at site $i$ and $j$, respectively. Here, $c_i^\dag$ ($c_i$) denotes the creation (annihilation) operator for a particle at site $i$. These particles can be either fermions or bosons. The third term describes the excitations of the crystal given by acoustic phonons, where  $a_{{\bi q},\lambda}^\dag$ ($a_{{\bf q},\lambda}$) creates (destroys) a phonon with quasimomentum ${\bi q}$ in the mode $\lambda$, with dispersion relation  $\omega_{{\bf q},\lambda}$. The last term in \eref{eq:1} is the particle-phonon coupling, which is of the density-displacement type, with coupling constant $M_{{\bf q},\lambda}$. The microscopic derivation of \eref{eq:1} is detailed in~\ref{AppA} and~\ref{AppB}. For the models we consider (see Section~\ref{settingup} and Section~\ref{rdc} below), we find that:

\begin{enumerate}

\item The phonon coupling  can largely exceed the hopping rate $J$, which precludes a naive treatment of the particle-phonon coupling as a (small) perturbation;

\item We are generally interested in the so-called {\it non-adiabatic regime}, where the
characteristic phonon frequency $\hbar \omega_{\rm D}$ (the {\it Debye frequency})
is typically (much) larger than the average kinetic energy of the particles $\sim J$, that is
$\hbar \omega_{\rm D} \gg J$, (see \Sref{rdc}). 
This is due to the fact that, unlike e.g. the case of electrons in ionic crystals,
in our system the mass of the particles and of the crystal dipoles are comparable,
and the crystals can be made stiff.
\end{enumerate}

Below, we derive a {\it master equation} for the dynamics of the particles only,
while the crystal phonons are treated as a thermal heat bath. The master equation in the Markovian limit has the general form
\begin{equation}
\dot{\rho}_{\rm S}(t)=-\frac{i}{\hbar}\big[ H_{\rm S},\rho_{\rm S}(t)\big] + \mathcal{D}[\rho_{\rm S}(t)],\label{eq:MasterGen}
\end{equation}
where $ H_{S}$ denotes a Hamiltonian term for the coherent time evolution of the reduced density matrix $\rho_{\rm S}$ for the particles, while $\mathcal{D}$ describes dissipative processes responsible for the incoherent dynamics. The coherent processes correspond to hopping of a particle dressed by the crystal phonons (a polaron) and polaron-polaron interactions, while the latter are thermally activated incoherent hopping, where a particle loses its phase coherence by emitting or absorbing a phonon in the hopping process. As suggested by points {\it i)} and {\it ii)}
above, we derive~\eref{eq:MasterGen} in a perturbative, strong-coupling, approach where the hopping rate $J$ in~\eref{eq:1} acts as the small parameter. In addition, we work in the {\it Brownian motion limit} of the master equation, where the characteristic time evolution of the system $\tau_{\rm S} \sim \max(1/J,1/V_{ij})$ is (much) slower than the characteristic time evolution of the thermal heat bath $\tau_{\rm B} \sim 1/\hbar \omega_{\rm D}$. We obtain explicit expressions for the coherent and incoherent (dissipative) contributions to the system time evolution, and show that the latter can be made negligible in a wide range of realistic parameters for our models.

We find that the effective Hamiltonian for the coherent dynamics of the particles only has the form
\begin{equation}
H_{\rm S}=-\tilde{J}\sum_{\langle i,j\rangle}c_i^{\dag}c_j+\frac{1}{2}\sum_{i,j}\tilde{V}_{ij}c_i^{\dag}c_j^{\dag}c_jc_i,\label{heff0}
\end{equation}
which corresponds to an extended Hubbard model. Here $\tilde{J}$ and $\tilde{V}_{ij}$ are the tunneling rate for the particles
and their mutual interactions, respectively, which are modified with respect to their bare values in~\eref{eq:1} by the coupling to the crystal phonons.

\subsection{Lang-Firsov Transformation (Polaronic picture)}\label{sec:LangFirsov}
In our system (see Section~\ref{rdc}), we find that particles are slow and strongly coupled to the crystal phonons, with Hamiltonian~\eref{eq:1}. 
Following Lang and Firsov~\cite{Alexandrov}, it is convenient to change from this picture of bare particles strongly coupled to phonons
to an equivalent one, where particles freely hop in the lattice while
carrying the lattice distortion. This corresponds to dressing the particles with the lattice phonons,
and the dressed particles are named {\it polarons}~\cite{Mahan}.
This is achieved by performing a unitary transformation of $H$ as $\bar H=U H U^\dag$ with
\begin{equation}\nonumber
U=\exp[\sum_{{\bf q},\lambda,j}u_{{\bf q},\lambda}e^{i{\bf q}{\bf r}^0_j}c_j^{\dag}c_j(a_{-{\bf q},\lambda}^{\dag}-a_{{\bf q},\lambda})]
\end{equation}
and $u_{{\bf q},\lambda}=M_{{\bf q},\lambda}/\hbar\omega_{{\bf q},\lambda}$ displacement amplitudes. In the new picture the Hamiltonian \eref{eq:1} reads~\cite{Mahan,Alexandrov}
\begin{eqnarray}
\bar{H}=-J\sum_{\langle i,j\rangle}c_i^{\dag}c_jX_i^{\dag}X_j - E_{\rm p}N_{\rm p}+\sum_{{\bf q},\lambda}\hbar\omega_{{\bf q},\lambda} &a_{{\bf q},\lambda}^{\dag}a_{{\bf q},\lambda}\nonumber\\
 &+ \frac{1}{2}\sum_{i,j}\tilde{V}_{ij}c_i^{\dag}c_j^{\dag}c_jc_i,\label{hlf}
\end{eqnarray}
where now $c_j$ ($c_j^\dag$) are the annihilation (creation) operators of a polaron located at site $j$. Here,
\begin{equation}
X_j=\exp[-\sum_{{\bf q},\lambda} u_{{\bf q},\lambda} e^{i{\bf q}{\bf r}_j^0}(a_{-{\bf q},\lambda}^{\dag}-a_{{\bf q},\lambda})].
\end{equation}
is the displacement operator for the crystal molecules due to the presence of a particle located at site $j$. The energy $E_{\rm p}$ in~\eref{hlf} is the {\it polaron shift}~\cite{Mahan}
\begin{equation}
E_{\rm p}\equiv\sum_{{\bf q},\lambda}M_{{\bf q},\lambda}^2/\hbar\omega_{{\bf q},\lambda}\label{eq:PolaronShift}
\end{equation}
and $N_{\rm p}\equiv \sum_j c_j^\dag c_j$ is the total number of particles. The quantity
\begin{equation}\label{eq:8}
\tilde{V}_{ij}= V_{ij}-2\sum_{{\bf q}}\frac{M_{{\bf q},\lambda}^2}{\hbar\omega_{{\bf q},\lambda}}\cos[{\bf q}({\bf r}_i^0-{\bf r}_j^0)] \equiv V_{ij} + \tilde V_{ij}^{(1)}
\end{equation}
is a modified particle-particle interaction, which comprises two terms: The first is the original (bare) interaction, while the latter is the {\it phonon mediated particle-particle interaction} $\tilde V_{ij}^{(1)}$, which in general (i) is long-ranged and (ii) can be comparable in strength to the bare interactions $V_{ij}$, see Section \ref{rdc}.

For $J=0$ the new Hamiltonian \eref{hlf} is diagonal and describes interacting polarons
and independent phonons. The latter are vibrations of the
lattice molecules around new equilibrium positions with
unchanged frequencies.
For small finite $J$, we can treat the first term in \eref{hlf} perturbatively, which is the starting point for the master equation approach detailed in the following sections.

\subsection{Effective dynamics of polarons inside a thermal crystal}\label{thermalcrystal}

In this section we derive a master equation describing the effective dynamics of the polarons embedded in the crystal, which we assume to be in thermal equilibrium. That is, we consider the phonons to provide a heat bath at temperature $T$ with a  density matrix given by
\begin{equation}
\rho^0_{\rm B} = 
\prod_{{\bf q},\lambda}\bar{n}_{{\bf q},\lambda}(T) \exp\left(-\frac{\hbar\omega_{{\bf q},\lambda}}{k_{\rm B}T}a_{{\bf q},\lambda}^\dag a_{{\bf q},\lambda}\right).
\end{equation}
Here $\bar{n}_{{\bf q},\lambda}(T)$ denotes the (Bose-Einstein) phonon distribution at temperature $T$,
\[\bar{n}_{{\bf q},\lambda}(T)=\langle a_{\bf q,\lambda}^\dag a_{\bf q,\lambda}\rangle = \frac{1}{e^{-\hbar\omega_{{\bf q},\lambda}/k_{\rm B}T}-1},\]
where $\av{O} \equiv\tr_{\rm B}\{O \rho_{\rm B}^0\}$ is the thermal expectation value of the bath operator $O$, with $\tr_{\rm B}$ denoting the trace over the bath degrees of freedom.

We split the Hamiltonian \eref{hlf} into three parts as $\bar H= H_{\rm S} +  H_{\rm B} +  H_{\rm I}$ with
\numparts
\begin{eqnarray}
H_{\rm S}&=&-J\sum_{\langle i,j\rangle}c_i^{\dag}c_j\langle X_i^{\dag} X_j\rangle + \frac{1}{2}\sum_{i,j}\tilde{V}_{ij}c_i^{\dag}c_j^{\dag}c_jc_i\label{sysHam},\\
H_{\rm B}&=&\sum_{{\bf q},\lambda}\hbar\omega_{{\bf q},\lambda} a_{{\bf q},\lambda}^{\dag}a_{{\bf q},\lambda}\label{bathH},\\
H_{\rm I}&=&-J\sum_{\langle i,j\rangle}c_i^{\dag}c_j(X_i^{\dag}X_j-\langle X_i^{\dag}X_j\rangle).\label{intH}
\end{eqnarray}
\endnumparts
Here, $H_{\rm S}$ is the ``reduced system'' Hamiltonian describing the dynamics of polarons, with a hopping rate $J\langle X_i^{\dag}X_j\rangle$ and interactions $\tilde V_{ij}$ for two polarons at site $i$ and $j$. The Hamiltonian $H_{\rm B}$ is the Hamiltonian for the bath, and Hamiltonian $H_{\rm I}$ gives the interactions between the system and the bath.
In writing~\eref{sysHam}-~\eref{intH} we have conveniently added a term $-J\sum_{\langle i,j\rangle}c_i^{\dag}c_j\langle X_i^{\dag} X_j\rangle$ to $H_{\rm S}$ and subtracted it from $H_{\rm I}$. This ensures that the thermal average over the interaction Hamiltonian vanishes, $\av{H_{\rm I}}=0$ \cite{Carmichael}.

The expectation value $\av{X_i^\dag X_j}$ in equation\eref{sysHam}
reads (see \ref{correlations})
\begin{equation}
\fl\av{X_i^\dag X_j}=e^{-S_T} \mathwith
S_T=2\sum_{{\bf q},\lambda}u_{{\bf q},\lambda}^2\sin^2\left[\frac{{\bf q}({\bf r}_i^0-{\bf r}_j^0)}{2}\right]\left(2\bar{n}_{{\bf q},\lambda}(T)+1\right).\label{st2}
\end{equation}
Equation~\eref{sysHam} shows that
\[ \tilde J\equiv J\av{X_i^\dag X_j} = Je^{-S_T}\]
plays the role of a phonon-modified tunneling rate, which is exponentially suppressed for $S_T > 0$. In the following we will often distinguish between two regimes, i.e. a weak coupling regime
where $S_T\ll 1$ (and $\tilde J \simeq J$) and a strong coupling regime where $S_T\gg 1$
(and $\tilde J \ll J$).


%


\subsubsection{The Quantum Brownian Motion Master Equation (QBMME)}\label{QBMME}
In this section we derive a master equation for the coherent and incoherent dynamics of the reduced system of interacting polarons in the Brownian motion limit, where the system time evolution, $\tau_{\rm S}$, is slow compared to the characteristic evolution time of the bath, $\tau_{\rm B}$. That this approach may provide a reasonable description of the system is suggested by the fact that in a wide range of parameters for atoms and molecules trapped in dipolar crystals we find $\tau_{\rm S} \sim \max(1/\tilde{J},1/\tilde{V}_{ij})\gg \tau_{\rm B} \sim 1/\omega_{\rm D}$. We will thus derive the coherent time evolution~\eref{heff0}, and provide explicit analytic expressions for the corrections both to the coherent and incoherent dynamics. In \Sref{rdc} we show that for the models of interest these corrections are in fact negligible in a wide regime of parameters, which provides an {\it a posteriori} self-consistency check for the approximations made here.\\

The time evolution for the density matrix of the entire system $\rho(t)$ comprising the (polaronic) reduced system and the heat bath is dictated by the Liouville-von Neumann equation
\[\dot{\tilde \rho}(t) = -\frac{i}{\hbar} \left[\tilde{H}_{\rm I}(t),\tilde \rho(t)\right],\]
where $\tilde{A}(t) \equiv e^{iH_0t/\hbar}Ae^{-iH_0t/\hbar}$ denotes an operator $A$ in the interaction picture with respect to $H_0=H_{\rm S}+H_{\rm B}$.

We assume that the reduced system and the heat bath are uncoupled at time $t=0$, so that $\rho(t=0)$  can be written as the tensor product $\rho(0)=\rho_{\rm S}(0)\otimes\rho^0_{\rm B}$, with $\rho_{\rm S}(0)$ and $ \rho^0_{\rm B}$ the density matrices of the reduced system and the bath, respectively. The condition $J \ll \hbar\omega_{\rm D}$, which states that the interaction is a weak perturbation,  forms the basis for a Born-Markov approximation with the phonons a finite temperature heat bath, providing the following master equation for the reduced density operator of the polarons
\begin{eqnarray}
\dot{\tilde\rho}_{\rm S}(t)&\approx& -\frac{1}{\hbar^2}\int_0^\infty d\tau\tr_{\rm B}\{[\tilde{H}_{\rm I}(t),[\tilde{H}_{\rm I}(t-\tau),\tilde{\rho}_{\rm S}(t)\otimes\rho_{\rm B}^0]]\}\\ 
&=&-\frac{J^2}{\hbar^2}\int_0^\infty\hspace{-.2cm} d\tau\hspace{-.3cm}\sum_{\langle i,j\rangle,\langle k,l\rangle}\hspace{-.2cm}\Big(\xi_{ij}^{kl}(\tau,T)\tilde{c}_{i}^\dag {(t)}\tilde{c}_{j}(t)\big{[}\tilde{c}_{k}^\dag(t-\tau)\tilde{c}_{l}(t-\tau),\tilde{\rho}_{\rm S}(t)]\nonumber\\
&&-\xi_{kl}^{ij}(-\tau,T)\big[\tilde{c}_{k}^\dag(t-\tau)\tilde{c}_{l}(t-\tau),\tilde{\rho}_{\rm S}(t)\big]\tilde{c}_{i}^\dag {(t)}\tilde{c}_{j}(t)\Big),\label{masta01}
\end{eqnarray}
where $\xi_{ij}^{kl}(\tau,T)$ are the (complex) bath correlation functions,
\begin{eqnarray}
\fl\xi_{ij}^{kl}(\tau,T)=\langle \tilde{X}_i^{\dag}(t)\tilde{X}_j(t)\tilde{X}_k^{\dag}(t')\tilde{X}_l(t')\rangle - \langle \tilde{X}_i^{\dag}(t)\tilde{X}_j(t)\rangle\langle\tilde{X}_k^{\dag}(t')\tilde{X}_l(t')\rangle\mid_{t'=t-\tau}\nonumber\\
=\langle X_i^{\dag}X_j\tilde{X}_k^{\dag}(-\tau)\tilde{X}_l(-\tau)\rangle-e^{-2S_T}={\xi_{kl}^{ij}}(-\tau,T)^*,\label{corrr}
\end{eqnarray}
which are treated in detail in \Sref{correlasect} and \ref{correlations}.

Under the additional condition $\max(\tilde{J},\tilde{V}_{ij})\ll (\hbar \omega_{\rm D}, k_{\rm B} T$), we focus on the Brownian motion limit of \eref{masta01}, where the system, like the interaction, evolves on a timescale much slower than the bath, see e.g.~\cite{Breuer}. In this limit we can approximate the system operators in equation~\eref{masta01} by
\begin{equation}
\tilde{c}_j(-\tau) \approx c_j-\frac{i}{\hbar}[H_{\rm S}, c_j]\tau,\label{eq:OpExpansion}
\end{equation}
and the master equation takes the form
\begin{eqnarray}
&\fl\dot{\rho}_{\rm S}(t)=-\frac{i}{\hbar}\Big[H_{\rm S}-\sum_{\langle i,j\rangle \langle k,l\rangle}\Delta_{ij}^{kl}(T)c_{i}^\dag c_{j}c_{k}^\dag c_{l},\rho_{\rm S}(t)\Big]\nonumber\\
&-\sum_{\langle i,j\rangle \langle k,l\rangle}\Gamma_{ij}^{kl}(T)\big(c_{i}^\dag c_{j}c_{k}^\dag c_{l}\rho_{\rm S}(t)+\rho_{\rm S}(t)c_{i}^\dag c_{j}c_{k}^\dag c_{l}-2c_{k}^\dag c_{l}\rho_{\rm S}(t)c_{i}^\dag c_{j}\big)\nonumber\\
&\fl-\frac{i}{\hbar}\hspace{-.2cm}\sum_{\langle i,j\rangle \langle k,l\rangle}\hspace{-.2cm}\gamma_{ij}^{kl}(T)\Big[c_{i}^\dag c_{j},
\Big[\tilde{J}\Big(\sum_{{k'}}c_{k'}^\dag c_{l}-\sum_{l'}c_{k}^\dag c_{l'}\Big)+\sum_m (V_{ml}-V_{mk})c_{k}^\dag c_{m}c_{m}^\dag c_{l},\rho_{\rm S}(t)\Big]\Big]\nonumber\\
&\fl-\frac{1}{\hbar}\hspace{-.2cm}\sum_{\langle i,j\rangle \langle k,l\rangle}\hspace{-.3cm}\delta_{ij}^{kl}(T)\Big[c_{i}^\dag c_{j},
\Big[\tilde{J}\Big(\sum_{{k'}}c_{k'}^\dag c_{l}\hspace{-.05cm}-\hspace{-.05cm}\sum_{l'}c_{k}^\dag c_{l'}\Big)\hspace{-.05cm}+\hspace{-.05cm}\sum_m (V_{ml}\hspace{-.05cm}-\hspace{-.05cm}V_{mk})c_{k}^\dag c_{m}c_{m}^\dag c_{l},\rho_{\rm S}(t)\Big]\Big],\label{mastaend}
\end{eqnarray}
with the quantities $\Delta_{ij}^{kl}(T),\Gamma_{ij}^{kl}(T),\delta_{ij}^{kl}(T)$ and $\gamma_{ij}^{kl}(T)$ given by
\numparts
\begin{eqnarray}
\Delta_{ij}^{kl}(T)&=&\frac{J^2}{\hbar}\int_0^\infty d\tau \Im[\xi_{ij}^{kl}(\tau,T)],\label{Delta}\\
\Gamma_{ij}^{kl}(T)&=&\frac{J^2}{\hbar^2}\int_0^\infty d\tau \Re[\xi_{ij}^{kl}(\tau,T)],\label{Gamma}\\
\gamma_{ij}^{kl}(T)&=&\frac{J^2}{\hbar^2}\int_0^\infty d\tau \tau\Re[\xi_{ij}^{kl}(\tau,T)],\label{gamma}\\
\delta_{ij}^{kl}(T)&=&\frac{J^2}{\hbar^2}\int_0^\infty d\tau \tau\Im[\xi_{ij}^{kl}(\tau,T)].\label{delta}
\end{eqnarray}
\endnumparts
Here $\Re[f(x)]$ and $\Im[f(x)]$ denote real and imaginary part of $f(x)$.


The first term inside the commutator on the r.h.s. of~\eref{mastaend} describes the coherent time evolution for the reduced system, with $H_{\rm S}$ the effective Hubbard Hamiltonian \eref{sysHam}. The terms proportional to $\Delta_{ij}^{kl}(T)$ in~\eref{mastaend} are self-energies, which in the single-polaron limit provide both a shift to the ground-state energy, and next-nearest-neighbor hopping terms~\cite{Yarlagadda05}. In addition, in the many-polaron problem they can provide offsite polaron-polaron interactions, whose strength will be evaluated in Section~\ref{sedrscwc} below. The terms on the r.h.s. of \eref{mastaend} proportional to $\Gamma_{ij}^{kl}(T)$ are related to incoherent hopping events where the number of phonons changes in the hop. 
That is, the polaron loses its phase coherence via the emission or absorption of phonons. These processes are thermally activated and can dominate over the coherent hopping rate $\tilde J$ for large enough temperatures $T$~\cite{Mahan}.

The terms proportional to $\gamma_{ij}^{kl}(T)$ and $\delta_{ij}^{kl}(T)$ are (small) corrections to the coherent and incoherent time evolution, respectively, which derive from the term proportional to $\tau$ in the expansion \eref{eq:OpExpansion}, and thus correspond to the Brownian motion  corrections to the time evolution of the reduced system.

We notice that the terms proportional to \eref{Delta}-\eref{delta} in~\eref{mastaend} do not identify
directly the corrections to the coherent time evolution given by~\eref{sysHam}, because equation~\eref{mastaend} is not diagonal. Instead these terms are elements of matrices whose eigenvalues are the corrections. The diagonalization of the master equation can be done analytically in the single polaron limit and is shown below in Section~\ref{diagonalizationn}.

In \Sref{sedrscwc} below we provide analytic expressions for the coherent and incoherent corrections to the coherent time evolution given by $H_{\rm S}$, cf.~\eref{sysHam}. In Section~\ref{rdc} we show that these corrections are in fact negligible in a wide range of realistic parameters for atoms and molecules, and
thus~$H_{\rm S}$ 
properly describes the dynamics of polarons inside the dipolar crystal.

\subsubsection{Correlation Functions:}\label{correlasect}
The bath correlation functions $\xi_{ij}^{kl}(\tau,T)$ appearing in the master equation~\eref{mastaend} are
computed in~\ref{correlations} and read
\begin{eqnarray}
\xi_{ij}^{kl}(\tau,T)=e^{-2S_T}\big(e^{-\Phi_{ij}^{kl}(\tau,T)}-1\big),\label{corro}\\
\fl\hspace{1.1cm}\mathwith\Phi_{ij}^{kl}(\tau,T)=\int dwJ_{ij}^{kl}(w) \Big[\coth\left( \frac{\hbar w}{k_{\rm B}T}\right)\cos (w\tau)-i\sin (w\tau)\Big].\label{phhii}
\end{eqnarray}
Here we have introduced the spectral density
\begin{equation}\label{specda}
J_{ij}^{kl}(w)=V_{\rm BZ}\sum_\lambda\int d{\bf q}^{d-1} \left[\frac{\partial \omega_{{\bf q},\lambda}}{\partial q_x}\right]^{-1} u_{{\bf q},\lambda}^2\bar{g}_{ij}^{kl}({\bf q})\Big|_{q_x(q^{d-1},w)},
\end{equation}
where $V_{\rm BZ}$ is the volume of the Brillouin zone, $q^{d-1}$ denotes all components
of the quasi-momentum vector except $q_x$, and $\bar{g}_{ij}^{kl}({\bf q})$ reads
\begin{eqnarray}
\bar{g}_{ij}^{kl}({\bf q})=&\cos[{\bf q}|{\bf r}_i^0-{\bf r}_k^0|]-\cos[{\bf q}|{\bf r}_j^0-{\bf r}_k^0|]\nonumber\\
&\hspace{3cm}-\cos[{\bf q}|{\bf r}_i^0-{\bf r}_l^0|]+\cos[{\bf q}|{\bf r}_j^0-{\bf r}_l^0|].
\end{eqnarray}
The quantity $\Phi_{ij}^{kl}$ is a decaying function of the time $\tau$
with $\max(|\Phi_{ij}^{kl}|)=2S_T$, the actual decay rate depending
strongly on the spectral density of the model~\cite{Mahan}.
In Sections \ref{sedrscwc} below we show
that at small finite temperatures this decay rate is fast enough to ensure that the corrections to the coherent time evolution in the QBMME~\eref{mastaend} for our 1D models are finite and small. This provides for an {\it a posteriori} check of the applicability of the QBMME to the polaron problem.

\subsubsection{Self energies and dissipation rates in the strong and weak coupling limits}\label{sedrscwc}
In this subsection we provide analytical approximate expressions for the matrix elements $\Delta_{ij}^{kl}$, $\Gamma_{ij}^{kl}$, $\gamma_{ij}^{kl}$ and $\delta_{ij}^{kl}$ in the limits of strong and weak
coupling $S_T \gg 1$ and $S_T \ll 1$, respectively. The details of the performed approximations are given in \ref{AppC}, while explicit results for our 1D models are shown below in Section~\ref{rdc}. Here we concentrate in particular on the leading self-energy corrections to the coherent time evolution determined by $H_S$ in~\eref{sysHam} in the two regimes.

In accordance with literature~\cite{Alexandrov}, we find that in the strong coupling limit, $S_T \gg 1$, the self-energies are suppressed by a factor of the order of $(J/E_{\rm p})^2$, with $E_{\rm p}$ the polaron shift~\eref{eq:PolaronShift}. In addition we find that in the weak coupling limit, $S_T \ll 1$, these corrections are strongly suppressed by a factor $(J/\hbar \omega_{\rm D})^2$ and as a consequence, in Section~\ref{rdc} we show that they are negligible in a wide range of realistic parameters for our models.\\

It is shown in~\ref{AppC} that all matrix elements $\Delta_{ij}^{kl}$, $\Gamma_{ij}^{kl}$, $\gamma_{ij}^{kl}$ and $\delta_{ij}^{kl}$ in the strong coupling limit are strongly suppressed by an exponential factor of the order $e^{-2S_T}$, unless $i=l$ and $j=k$,~\cite{Mahan,Alexandrov}. This makes sense, since processes for which $i\neq l$ and $j \neq k$ correspond to a double hop of a polaron, each one being suppressed by the factor $e^{-S_T}$. The conditions $i=l$ and $j=k$ describe a "swap"-process, which corresponds, e.g. in the case of $\Delta_{ij}^{kl}$, to the virtual hop of a polaron from its current position to a neighboring site and back.

The various corrections read
\numparts
\begin{eqnarray}
\Delta_{01}^{10}(T) &\approx&\frac{J^2}{\hbar \omega_{\rm D}}\frac{\pi^{1/2}}{2}\frac{e^{-B^2/4A_T}}{\sqrt{A_T}}{\rm Erfi}(B/2\sqrt{A_T}),\label{scDelta}\\
\Gamma_{01}^{10}(T) &\approx&\frac{J^2}{\hbar^2 \omega_{\rm D}}\frac{\pi^{1/2}}{2}\frac{e^{-B^2/4A_T}}{\sqrt{A_T}},\label{scGamma}\\
\gamma_{01}^{10}(T) &\approx&\frac{J^2}{\hbar^2 \omega_{\rm D}^2}\left(\frac{1}{2A_T}-\frac{\pi^{1/2}}{4}\frac{B e^{-B^2/4A_T} {\rm Erfi}(B/2\sqrt{A_T})}{ A_T^{3/2}}\right),\label{scgamma}\\
\delta_{01}^{10}(T) &\approx&\frac{J^2}{\hbar^2 \omega_{\rm D}^2}\frac{\pi^{1/2}}{4}\frac{B e^{-B^2/4A_T}}{ A_T^{3/2}},\label{scdelta}
\end{eqnarray}
\endnumparts
where ${\rm Erfi}$ denotes the error function and where we have introduced the quantities
\begin{eqnarray}
A_T&\equiv&\int dw\frac{1}{2}J_{01}^{10}(w)w^2\coth\left(\frac{\hbar w}{k_{\rm B}T}\right),\\
B&\equiv&\int dw J_{01}^{10}(w)w.
\end{eqnarray}
Equations \eref{scDelta}-\eref{scdelta} result from an expansion of the function $-\Phi_{ij}^{kl}(\tau,T)$ appearing in the exponent of $\xi_{ij}^{kl}(\tau,T)$, cf. \eref{corro}, up to second order in the time $\tau$ around its maximum.
For a vanishing $A_T$, which corresponds to neglecting second order terms in the expansion of $-\Phi_{ij}^{kl}(\tau,T)$, we find
\begin{eqnarray}
\Delta_{01}^{10} &= \frac{J^2}{\hbar \omega_{\rm D} B},\\
\gamma_{01}^{10} &= \frac{J^2}{(\hbar \omega_{\rm D} B)^2}.
\end{eqnarray}
A simple estimate of $B$, see~\ref{anaResults}, shows that it is of the order of $B \sim 2E_{\rm p}/\hbar \omega_{\rm D}$ for a sufficiently strong coupling and therefore $\Delta_{01}^{10}/E_{\rm p} \propto (J/E_{\rm p})^2$ and $\gamma_{01}^{10} \propto (J/E_{\rm p})^2$. This dependence is known in literature as the ``$1/\lambda$'' strong coupling expansion, where $\lambda = E_{\rm p}/J$, see ~\cite{Alexandrov}.

In the weak coupling limit, $S_T \ll 1$,  the correlation functions $\xi_{ij}^{kl}$ can be expanded to first order in $\Phi_{ij}^{kl}(\tau,T)$, which leads to the following expressions for the matrix elements
\numparts
\begin{eqnarray}
\Delta_{ij}^{kl}(T)&\approx&\frac{J^2}{\hbar \omega_{\rm D}}e^{-2S_T}P\hspace{-.2cm}\int dw \frac{J_{ij}^{kl}(w)}{w}\label{eq:Deltawc}\\
\Gamma_{ij}^{kl}(T) &\approx& \frac{J^2}{\hbar^2 \omega_{\rm D}} e^{-2S_T}\pi\lim_{w\rightarrow 0}J_{ij}^{kl}(w)\coth\left(\frac{\hbar w}{k_{\rm B}T}\right)\label{eq:Gammawc}\\
\gamma_{ij}^{kl}(T) &\approx& -\frac{J^2}{\hbar^2 \omega_{\rm D}^2} e^{-2S_T}P\hspace{-.2cm}\int dw \frac{J_{ij}^{kl}(w)}{w^2}\coth\left(\frac{\hbar w}{k_{\rm B}T}\right)\label{eq:gammawc}\\
\delta_{ij}^{kl}(T)&\approx&-\frac{J^2}{\hbar^2 \omega_{\rm D}^2}e^{-2S_T}\pi\partial_{w}J_{ij}^{kl}(0),\label{eq:deltawc}
\end{eqnarray}
\endnumparts
as detailed in~\ref{anaResults}. Here, $P\hspace{-.15cm}\int dx$ denotes the Cauchy principal value integral. In contrast to the strong coupling regime, as shown in~\ref{AppC} here the (small) parameter that gives the approximate size of the corrections is  $J/\hbar \omega_{\rm D}$ and not $J/E_{\rm p}$. Explicit results for the corrections in this limit are given in Section~\ref{rdc} below. The ratio $(J/\hbar \omega_{\rm D})^2$ also determines the size of the incoherent processes $\hbar\Gamma_{ij}^{kl}$.
This can be seen by performing a low temperature approximation of expression~\eref{eq:Gammawc} above, for which we find $\hbar\Gamma_{ij}^{kl}(T)\propto \tilde{J}^2 k_{\rm B}T/(\hbar\omega_{\rm D})^2$. We notice that, in addition to being proportional to the small factor $(J/\hbar\omega_{\rm D})^2$, these corrections depend linearly on temperature, a result also found in~\cite{Mahan}. Because dipolar crystals can have a large Debye frequency, in our models we find $J/\hbar \omega_{\rm D} \ll 1$ and thus corrections to the coherent time evolution determined by $H_S$ are small.

\subsubsection{Diagonalization of the master equation}\label{diagonalizationn}
As noted above the corrections to the master equation $\Gamma_{ij}^{kl}(T), \Delta_{ij}^{kl}(T), \gamma_{ij}^{kl}(T)$ and $\delta_{ij}^{kl}(T)$ are just matrix elements, while the actual
energies and rates are obtained by diagonalizing the various terms in~\eref{mastaend}. In the following we sketch how to perform this diagonalization in the simple case of a single polaron for
$\Delta_{ij}^{kl}$ and $\Gamma_{ij}^{kl}$, while similar computations for all coherent and incoherent corrections are shown in \ref{diagonalization}. In the case of a single polaron the terms proportional
to $\Delta_{ij}^{kl}$ are easily diagonalized, since the eigenergies for a particle in a lattice
are readily determined by the energies associated with the various quasimomenta. That is

\begin{eqnarray}
\sum_{\langle ij\rangle \langle kl\rangle}\Delta_{ij}^{kl}(T)c_{i}^\dag c_j c_{k}^\dag c_{l} &= \sum_{i}\sum_{m,n}\Delta_{i,i+m}^{i+m,i+m+n}c_i^\dag c_{i+m+n}\\
&=\sum_{\bf q}\sum_{m,n}\Delta_{0,m}^{m,m+n}e^{i {\bf q}({\bf r}_m^0+{\bf r}_n^0)}c_{\bf q}^\dag c_{\bf q}\label{eq:c40}
\end{eqnarray}
where the sums over $m,n$ range over basis vectors in the lattice,
and  $\sum_i c_i^\dag c_{i+m}=\sum_{\bf q} e^{i {\bf q}{\bf r}_m^0}c_{\bf q}^\dag c_{\bf q}$. The
eigenvalues can now be directly read-off from~\eref{eq:c40}, as
$\Delta_{{\bf q}}=\sum_{m,n}\Delta_{0,m}^{m,m+n}e^{i {\bf q}({\bf r}_m^0+{\bf r}_n^0)}$.
The largest eigenvalue gives an upper bound to the energy of this term in the master equation.

To determine the rate of the dissipative term in~\eref{mastaend}
we notice that this term can be written as
\begin{equation}
\frac{1}{\hbar}\sum_{\langle ij\rangle\langle kl\rangle}\hbar\Gamma_{ij}^{kl}(T)\big( \{c_{i}^\dag c_j c_{k}^\dag c_{l},\rho_{\rm s}(t)\} - 2 c_{k}^\dag c_{l}\rho_{\rm s}(t)c_{i}^\dag c_{j}\big).
\end{equation}
The amplitude of the dominant rate can be now estimated from  $\sum_{\langle ij\rangle\langle kl\rangle}\hbar\Gamma_{ij}^{kl}(T) c_{i}^\dag c_j c_{k}^\dag c_{l}$, whose eigenvalues read
\begin{equation}
\Gamma_{{\bf q}}=\sum_{m,n}\Gamma_{0,m}^{m,m+n}e^{i {\bf q}({\bf r}_m^0+{\bf r}_n^0)}.
\end{equation}
Calculations similar to the ones above lead to the eigenvalues $\gamma_{\bf q}$ and $\delta_{\bf q}$ for the Brownian motion corrections (see \ref{diagonalization}).

In one dimension, which is relevant for the models of Section~\ref{rdc} below, we find
\begin{eqnarray}
\Delta_q(T) &=  2 [\Delta_{01}^{10}(T)+\Delta_{01}^{12}(T)\cos(qa)],\label{eq:Delta1D}\\
\Gamma_{q}(T) &= 2 [\Gamma_{01}^{10}(T)+\Gamma_{01}^{12}(T)\cos(qa)],\\
\gamma_q(T) &= 2[(\gamma_{01}^{23}(T)-\gamma_{01}^{13}(T))\cos(3qa)\nonumber\\
&\hspace{.5cm}+(\gamma_{01}^{21}(T)+\gamma_{01}^{01}(T)+\gamma_{01}^{0,-1}(T)-\gamma_{01}^{1,-1}(T))\cos(qa)],\\
\delta_q(T) &= 2[(\delta_{01}^{23}(T)-\delta_{01}^{13}(T))\cos(3qa)\nonumber\\
&\hspace{.5cm}+(\delta_{01}^{21}(T)+\delta_{01}^{01}(T)+\delta_{01}^{0,-1}(T)-\delta_{01}^{1,-1}(T))\cos(qa)].\label{eq:delta1D}
\end{eqnarray}\\

In conclusion, we note that the corrections to the coherent time evolution given by
$\Delta_{ij}^{kl}$, $\Gamma_{ij}^{kl}$, $\gamma_{ij}^{kl}$ and $\delta_{ij}^{kl}$ can
be made small both in the strong {\it and} weak coupling regimes, by ensuring
that the ratios $J/ E_{\rm p}$ and $J/\hbar \omega_{\rm D}$ are small, respectively.
The smallness of these corrections provides for an {\it a posteriori} check of the
applicability of the Brownian motion master equation approach to the polaron problem.

\section{Crystals of polar molecules}\label{settingup}
\begin{figure*}[t]
\begin{flushright}
\includegraphics[width=.9\columnwidth]{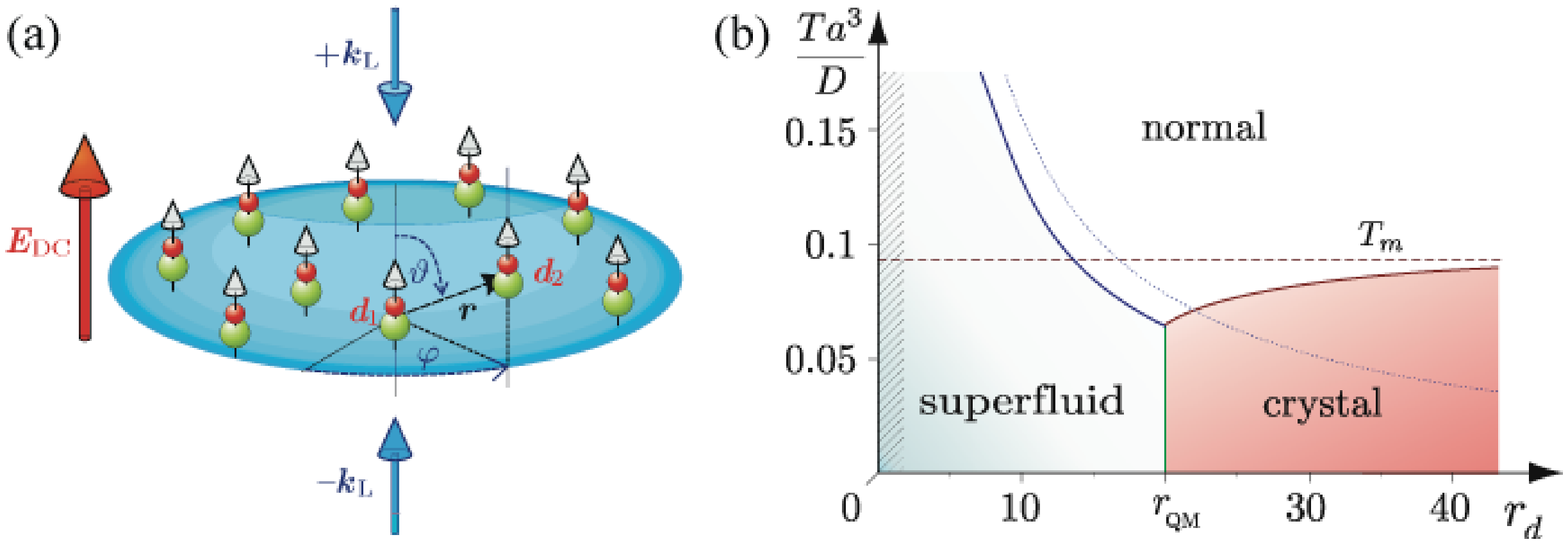}
\end{flushright}
\caption{\label{figSphase} \label{fig:fig1}(a) System setup: Polar molecules are
    trapped in the ($x,y$)-plane by an optical lattice made of two
    counter-propagating laser beams with wavevectors $\pm{\bf k}_{\rm
      L}=\pm{k}_{\rm L}{\bf e}_z$, (blue arrows). The dipoles are
    aligned in the $z$-direction by a DC electric field ${\bi E}_{\rm
      DC}\equiv E_{\rm DC}{\bi e}_z$ (red arrow). (b) Phase diagram in the $T-r_{d}$ plane:
crystalline phase for interactions
$r_{d}>r_{\mathrm{\scriptscriptstyle QM}}$ and temperatures below
the classical melting temperature $T_{m}$ (dashed line)~\cite{Kalia}. The crossover to the
unstable regime for small replusion and finite confinement
$\omega_{\perp}$ is indicated (hatched region).}
\end{figure*}

In this section we briefly review how to realize self-assembled crystals of polar molecules. Following \cite{BuechlerPRL2007,PupilloPRL2008}, here we focus on crystals in two and one dimensions. However, self-assembled crystals in three-dimensions can also be realized as detailed in~\cite{GorshkovPRL08}.

{\it Two-dimensional crystals:} We consider a system of cold polar molecules in the presence of a DC electric
field under strong transverse confinement, as illustrated in \fref{fig:fig1}(a).
A weak DC field along the $z$-direction induces a dipole moment $d_{\rm c}$ in the ground state
of each molecule. Thus, the molecules interact via the
dipole-dipole interaction $V_{{\rm dd}}^{{\rm 3D}}({\bf r})=D(r^{2}-3z^{2})/r^{5}$,
with $D=d_{\rm c}^{2}$. This interaction is purely repulsive for molecules confined
to the $x,y$-plane, while it is attractive for $z>r/\sqrt{3}$, leading
to an instability towards collapse in the many body system.
In reference~\cite{BuechlerPRL2007} it is shown that this instability can be suppressed
for a sufficiently strong 2D confinement along $z$,
as provided, for example, by a deep optical lattice with frequency
$\hbar \omega_\perp$ (blue arrows in the figure). In fact, a strong confinement
with $\hbar \omega_\perp \gtrsim D/a^3$, with $a$ the mean interparticle distance,
confines the molecules to distances larger than
\begin{equation}
a_{\rm min}=\left( \frac{12 d_{\rm c}^2}{m_{\rm c} \omega_\perp} \right)^{1/5},
\end{equation}
where $V_{{\rm dd}}^{{\rm 3D}}({\bf r})$ is purely repulsive,
and thus the system is collisionally stable. Here, $m_{\rm c}$ is the mass of the molecules.
The 2D dynamics in this pancake configuration is described by the
Hamiltonian
\begin{equation}
H_{{\rm dd}}^{{\rm 2D}}=\sum_{i}\frac{{\bf p}_{\mathbf{\rho}_i}^{2}}{2m}+\sum_{i<j}V_{{\rm dd}}^{{\rm 2D}}({\bf \rho}_{ij}),\label{hamilton1}
\end{equation}
which is obtained by integrating out the fast $z$-motion. Equation~(\ref{hamilton1})
is the sum of the 2D kinetic energy in the $x$,$y$-plane and an effective repulsive
2D dipolar interaction
\begin{eqnarray}
V_{{\rm dd}}^{{\rm 2D}}({\bf \rho})=D/\rho^{3},\label{eq:eqIn}
\end{eqnarray}
with ${\bf \rho}_{ij}\equiv(x_{j}-x_{i},y_{j}-y_{i})$ a vector in the
$x,y$-plane.
Tuning the induced dipole moment $d_{\rm c}$ drives the system
from a weakly interacting gas (a 2D superfluid in the case of bosons),
to a crystalline phase in the limit of strong repulsive dipole-dipole
interactions. This crystalline phase corresponds to the limit of strong repulsion where
particles undergo small oscillations around their equilibrium positions.
The strength of the interactions is characterized by the ratio $r_d$
of the interaction energy over the kinetic energy
at the mean interparticle distance $a$
\begin{eqnarray}
r_{d}\equiv\frac{E_{\rm pot}}{E_{\rm kin}}
=\frac{D/a^{3}}{\hbar^{2}/ma^{2}}=\frac{Dm}{\hbar^{2}a}.\label{eq:eqrd}
\end{eqnarray}
This parameter is tunable
as a function of $d_c$ from small to large $r_{d}$. A crystal forms for
\begin{equation}
r_{d} \geq r_{\rm c}=18 \pm 4,
\end{equation}
where the interactions are dominant~\cite{BuechlerPRL2007,Astrakharchik}.
For a dipolar crystal, this is the limit of large densities,
as opposed to Wigner crystals with $1/r$- Coulomb interactions.

\begin{figure}[t!]
\begin{flushright}
\includegraphics[width=.85\columnwidth]{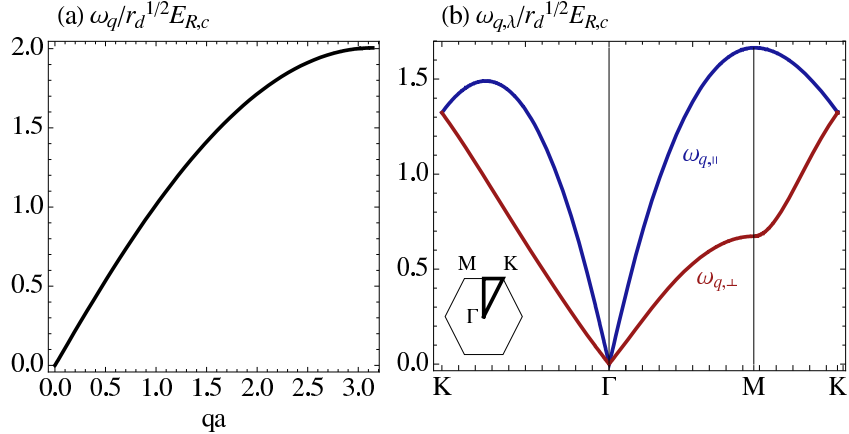}
\end{flushright}
\caption{\label{figdispersion} We show the dispersion for a 1D and 2D dipolar crystal as a function of the quasimomentum in units of the crystal recoil energy $E_{\rm R,c}$. In 1D (a) the dispersion is peaked at the zone border and tends to zero $\propto q$ for small momenta. The 2D dispersion has two acoustic branches, a longitudinal and a transversal one. We plot the two dispersions as a function of the quasimomentum, choosing a path in the first Brillouin zone that is outlined in the inset of the figure.}
\end{figure}

\Fref{figSphase}(b) shows a schematic phase diagram for a
dipolar gas of bosonic molecules in 2D as a function of $r_{d}$ and
temperature $T$. 
In the limit of strong interactions $r_{d}> r_{\rm c}$ the polar
molecules are in a crystalline phase for temperatures $T<T_{\rm m}$ with
$T_{\rm m}\approx0.09D/a^{3}\simeq 0.018r_d E_{{\rm R,c}}$,
with $E_{{\rm R,c}}\equiv \pi^2 \hbar^2/2 m a^2$~\cite{Kalia} the crystal recoil energy, typically a few to tens of kHz. The configuration with minimal energy is a triangular lattice with spacing $a_{\mathrm{\scriptscriptstyle L}} = (4/3)^{1/4} a$ see \cite{BuechlerPRL2007}. Excitations of the
crystal are acoustic phonons with Hamiltonian given by equation \eref{bathH}, and characteristic Debye frequency
$\hbar\omega_{{\rm D}}\sim1.6\sqrt{r_{d}}E_{{\rm R,c}}$.
The dispersion relation for the phonon excitations is obtained in~\ref{2Dphononspectrum} and shown in \fref{figdispersion}(b) below.\\

{\it One-dimensional crystals:} One dimensional analogues of the 2D crystals can
be realized by adding an additional {\em in-plane}
optical confinement to the configuration of \fref{figSphase}(a)~\cite{Citro,Rabl07,PupilloPRL2008}.
For large enough interactions $r_d \gg 1$, the phonon frequencies have the simple form
$\hbar\omega_{q}=(2/\pi^{2})\left[12r_{d}f_{q}\right]^{1/2}E_{{\rm
R,c}}$, with $f_{q}=\sum_{j>0}4\sin(qaj/2)^{2}/j^{5}$, see \fref{figdispersion}(a). The Debye
frequency is $\hbar\omega_{{\rm
D}}\equiv\hbar\omega_{\pi/a}\sim1.4\sqrt{r_{d}}E_{{\rm R,c}}$, while
the classical melting temperature can be estimated to be of the
order of $T_{\rm m}\simeq 0.2 r_d E_{{\rm R,c}}/k_{\rm B} $, see \cite{Rabl07}.\\

Finally, for a given induced dipole $d_{\rm c}$ the ground-state of an ensemble of
polar molecules is a crystal for mean interparticle distances
$a_{{\rm min}}\lesssim a \lesssim a_{\rm max}$, where
$a_{\rm max}\equiv d_{\rm c}^2 m/\hbar^2 r_{\rm c}$
corresponds to the distance at which the crystal melts into a
superfluid. For SrO (RbCs) molecules with the permanent dipole moment
$d_{\rm c}=8.9$D ($d_{\rm c}=1.25$D), $a_{\rm min}\sim 200$nm($100$nm), while
$a_{\rm max}$ can be several $\mu$m. Since for large enough
interactions the melting temperature $T_{\rm m}$ can be of order of
several $\mu$K, the self-assembled crystalline phase should be
accessible for reasonable experimental parameters using cold polar
molecules.

\section{Specific implementations with atoms and molecules in dipolar crystals}\label{rdc}

In this section we consider a mixture of two species of particles confined to one dimension. The first species of particles comprises (strongly interacting) molecules with dipole moment $d_{\rm c}$, forming a one-dimensional crystal. The second species of particles can be either atoms [see \fref{figs:fig1}(b)] or molecules of a second species, with dipole moment $d_{\rm p} \ll d_{\rm c}$ [see \fref{figs:fig1}(c)]. The former interact with the crystal molecules via a short range pseudopotential proportional to an elastic scattering
length $a_{{\rm cp}}$, while the latter interact with the crystal molecules via long-range dipole-dipole interactions. For both configurations, we obtain explicit expressions for all parameters characterizing the coherent and the incoherent dynamics in the system. These one dimensional setups can be readily generalized to two dimensions.

\subsection{Neutral atoms moving inside a crystal tube}

As a first realization, we consider a setup, where an ensemble of neutral atoms is confined by an optical trap to the same 1D tube as the dipolar crystal, see \Fref{figs:fig1}(b), say along ${\bi x}$. For simplicity we assume the trap for the neutral atoms and the polar molecules to have the same harmonic oscillator frequency $\omega_\perp$.

An atom and a molecule inside the tube interact via a short range potential, which we model  in the form of an effective 1D zero range potential,
\[ V_{\rm cp}(x-X) = g_{\rm cp}  \delta(x-X) \]
Assuming that the 3D scattering length $a_{\rm cp}$ is (much) smaller than the harmonic oscillator length of the traps, $a_{\rm cp}\ll a_{{\rm p},\perp}=(\hbar/m_{\rm p}\omega_\perp)^{1/2}$, the effective 1D coupling strength is given by $g_{\rm cp}\approx2\hbar\omega_\perp a_{\rm cp}$. In the following we focus on positive scattering lengths, $a_{\rm cp}>0$, corresponding to a situation where the atoms and molecules effectively repel each other, cf. $g_{\rm cp}>0$.

\subsubsection{Tight binding limit and Hubbard models.}

For a ``frozen'' crystal, i.e. without phonons, the molecules are at their equilibirum positions, $ja+a/2$, which provide for a static periodic potential for the atoms,
\begin{equation}
 V_{\rm p}(x) = g_{\rm cp} \sum_j \delta(x-ja-a/2).
 \end{equation}

 \begin{figure}[t!]
\begin{flushright}
\includegraphics[width=.85\columnwidth]{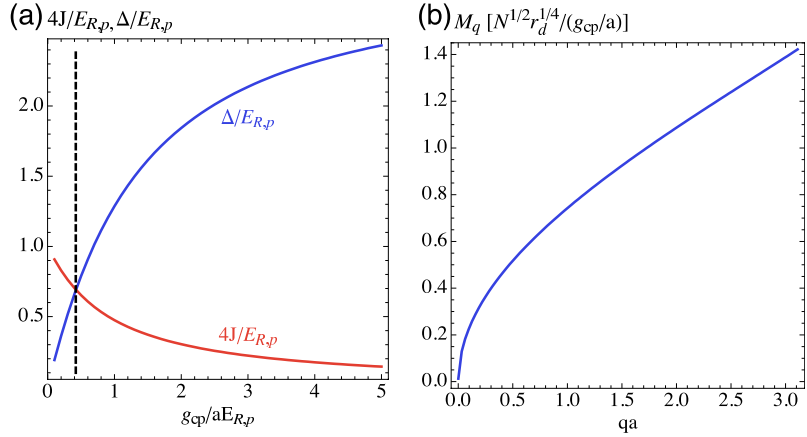}
\end{flushright}
\caption{\label{mod1fig1}
(a) The bandwidth $4J$ of the lowest band (solid blue line) and the gap $\Delta$ to the first excited band (solid red line) for a neutral atom scattered from a 1D potential comb of strength $g_{\rm cp}$ and lattice spacing $a$. All energies are given in terms of the particle recoil energy $E_{\rm R,p}$. The dashed line denotes the coupling strength, $g_{\rm cp}/a\approx E_{R,p}/2$, where the gap and band-width are equal. (b) The corresponding  particle phonon coupling $M_q$ in units of $r_d^{1/4}\sqrt{N}E_{\rm R,p}$ as a function of the quasi momentum $q$ of the atom. The interaction is linear and peaked for large $q$, while it shows a square-root behavior for small quasi momenta (see text).}
\end{figure}

The dynamics for a single neutral atom is then determined from the static Hamiltonian $H_{\rm p}=p^2/2m_{\rm p}+V_{\rm p}(x)$, corresponding to the Kronig-Penney model with a potential comb of strength $g_{\rm cp}$. Its energy spectrum is given in the form of a band-structure, $E_{n,q}$ (with band-index $n=0,1,\ldots$), which is obtained from
\begin{equation}
\frac{\pi}{4\alpha}\sum_\pm\cot\left(\frac{\pi \alpha\pm qa}{2}\right)=\frac{aE_{\rm R,p}}{g_{\rm cp}} \mathwith \alpha\equiv\left(\frac{E_{n,q}}{E_{\rm R,p}}\right)^{1/2}
\end{equation}
where $E_{\rm R,p}\equiv \hbar^2\pi^2/2m_{\rm p}a^2$ denotes the recoil energy of an atom. We denote the band-width of the lowest band by $4J \equiv E_{0,\pi/a}-E_{0,0}=E_{\rm R,p}-E_{0,\pi/a}$, and the gap to the first band by $\Delta\equiv E_{1,\pi/a}-E_{0,\pi/a}=E_{1,\pi/a}-E_{\rm R,p}$. The latter are both shown in \fref{mod1fig1}(a) as a function of the coupling strength $g_{\rm cp}$. We notice that for $g_{\rm cp}\gtrsim E_{\rm R,p}a/2$ (indicated by a vertical dashed line), the gap exceeds the band-width, $\Delta>4J$. In the tight binding limit, cf. $g_{\rm cp}\gg E_{\rm R,p}a$, the dispersion relation in the lowest band becomes $E_{0,q}\approx 4J\sin^2(qa/2)$, with $J\approx 2 E_{\rm R,p}^2/\pi^2g_{\rm cp}+\Or(aE_{\rm R,p}/g_{\rm cp})$ and $\Delta\approx3E_{\rm R,p}$. The Wannier-functions for a particle become localized at site $j$ and are approximated by $w_j(x)\approx \cos[\pi(x-ja)/a]/\sqrt{a/2}$ for $|x-ja|<a/2$ and zero otherwise. 

Obtaining the tight-binding limit requires that the ratio,
$g_{\rm cp}/aE_{\rm R,p}=2a_{\rm cp}a/\pi^2a_{{\rm p},\perp}^2$
(largely) exceed the value $\approx 1/2$. We notice that for current state-of the art optical traps, one can achieve harmonic oscillator lengths as small as $a_{{\rm p},\perp}\sim 20~{\rm nm}$, and taking a ``typical'' 3D scattering length of  $a_{\rm cp}\sim 100a_0\approx 5~{\rm nm}$, we get that $g_{\rm cp}/aE_{\rm R,p}\gtrsim1/2$ is attained for lattice spacings $a\gtrsim \pi^2a_\perp^2/a_{\rm cp}\sim 200{\rm nm}$.\\

Analogous to the atom-molecule interactions, we model the interactions between two neutral atoms by a contact potential with a coupling strength given by $g_{\rm pp}\approx \hbar\omega_\perp a_{\rm pp}$ for the 3D atom-atom scattering length $a_{\rm pp}\ll a_\perp$. Then in the tight-binding limit the atom-atom interactions reduce to repulsive onsite energy shifts only
\[
V_{i,j} = g_{\rm pp} \int dx w_i(x)^2 w_j(x)^2 \approx \frac{3}{2}\frac{g_{\rm pp}}{a} \delta_{i,j}.
\]
The dynamics for an ensemble of bosonic atoms in the crystal is then described by a single band Bose-Hubbard model with hopping rate $J$ and onsite repulsion $V_{ii}$, provided that $V_{ii}\ll\Delta$. On the other hand for an ensemble of (spin-polarized) fermionic atoms, we notice that the system reduces to a lattice model with hopping rate $J$ and no interactions.

The atoms couple to the crystal molecules via a density-displacement interaction~\cite{Mahan}. To first order in the displacement (see \ref{appCPinteraction}), the coupling constant reads
\begin{equation}\label{theMmod1}
M_q=\left(\frac{\hbar}{2N_{\rm c} m_{\rm c}\omega_{q}}\right)^{1/2}q\beta_{q}\tilde{V}_{\rm cp}(q) =
\frac{g_{\rm cp}}{a}\sqrt{\frac{2\hbar}{Nm_{\rm c}\omega_q}}|q|\beta_q ,
\end{equation}
where $\tilde{V}_{\rm cp}$ is the Fourier transform of the atom-molecule interaction potential, and $\beta_q$ is the Fourier transform of the square of the Wannier-functions. For $g_{\rm cp}\gg aE_{\rm R,p}$ the latter is
\[\beta_q = \int dx w_0(x)^2e^{iqx} \approx  \frac{8\pi^2\sin^2\frac{qa}{2}}{4\pi^2 qa - q^3a^3}.\]
The latter decreases with increasing $q$ from $\beta_0\equiv 1$ to $\beta_{\pi/a}=8/3\pi\approx0.85$. The (monotonical) dependence of the coupling constant $M_q$ on the quasi-momentum is shown in \fref{mod1fig1}(b). In particular, it has a maximum, $M_{\pi/2}\approx(8g_{\rm cp}/3a^2)(2\hbar/Nm_{\rm c}\omega_{\rm D})^{1/2}$, at the band-edges, while for small quasi-momenta $q$ it vanishes as $|q|^{1/2}$, i.e. $M_{q}\approx (2g_{\rm cp}/a^2)(\hbar|qa|/Nm_{\rm c}\omega_{\rm D})^{1/2}+ \Or(qa)^{5/2}$.

Finally, let us address the validity of the {\em single} band approximation in the Hubbard model \eref{eq:1}, when coupled to phonons. For simplicity let us consider the limit of vanishing interactions $V_{ii}=0$ and a weak coupling $M_q$:
We notice that, for $\hbar\omega_{\rm D} < 4J+\Delta$ the second band is gapped from the branch of acoustic phonons, and therefore higher band excitations are (strongly) suppressed and can be neglected. Since $E_{\rm R,p}\leq4J+\Delta\leq3E_{\rm R,p}$, this requires a mass ratio $m_{\rm p}/m_{\rm c}\lesssim 3/\sqrt{2r_d}$. While this for a soft crystal with $r_d\sim1$ merely implies that $m_{\rm p}<m_{\rm c}$, for a stiff crystal with $r_d\sim200$ this requires $m_{\rm p}\lesssim 0.15m_{\rm c}$, which is quite restrictive.
In the latter case, that is for a stiff crystal and comparable masses, cf. $\sqrt{2r_d}m_p/3m_{\rm c}>1$, we notice that the first excited band would cut the phonon branch at a frequency $\omega_\star \sim 4J + \Delta$. However, in this regime a single band model may still hold, if one restricts the initial phonon population to sufficiently low temperatures, i.e. for $k_{\rm B}T\ll \hbar\Delta$. In the tight binding limit this requires temperatures $T\ll (3m_c/\sqrt{2r_d}m_p)\times\hbar\omega_{\rm D}/k_{\rm B}$ which even for a stiff crystal with $r_d\sim200$ and comparable mass ratio $m_p\sim m_c$ yields temperatures on the order of $0.1\hbar\omega_{\rm D}/k_{\rm B}$. These are smaller than the melting temperature of the crystal, $T_{\rm m}\simeq 0.1\sqrt{2r_d}\hbar\omega_{\rm D}/k_{\rm B}$ (in 1D), and are thus reasonable, even for a ``soft'' crystal with $r_d\sim1$.

\subsubsection{Extended Hubbard model for atomic polarons inside a dipolar crystal}
As we have seen in \Sref{sec:LangFirsov}, it is convenient to change from a picture of bare atoms and crystal phonons, to one of atoms dressed by their surrounding crystal displacements, i.e. polarons.
The corresponding displacement amplitudes $u_q$ for the dressing then are %
\begin{equation}\label{eq:35}
u_{q}=\frac{M_q}{\hbar\omega_q}=\frac{1}{\sqrt{N}}\frac{2\pi^2U_0}{(186\zeta(5))^{3/4}}\beta_q\frac{qa}{{\rm w}_q^{3/2}},
\end{equation}
which at the band-edge attain their minimum, $u_{\pi/a}\approx 1.02 U_0^2/N^{1/2}$, while they diverge at small quasimomenta as $\sim 4.94 |qa|^{-1/2}$. In \eref{eq:35} we introduced the coupling ratio 
\[U_0=\frac{g_{\rm cp}}{aE_{\rm R,c}r_d^{3/4}}=\frac{g_{\rm cp}}{aE_{\rm R,p}} \frac{m_{\rm c}}{m_{\rm p}} \frac{1}{r_d^{3/4}},\]
which increases linearly with the atom-molecule coupling constant $g_{\rm cp}/aE_{\rm R,p}$ and the mass ratio of $m_c/m_p$, but is inversely proportional to the ``stiffness'' of the crystal $r_d$. The dependence of $U_0\times m_p/m_c=g_{\rm cp}/aE_{\rm cp}r_d^{3/4}$ on the coupling $g_{\rm cp}/aE_{\rm cp}$ and $r_d$ is also shown as dashed contour lines in \fref{mod1fig1_5}(a). We see that for, e.g., a mass ratio of $m_p/m_c\sim1$ (e.g. for a gas of Cs atoms in a crystal of LiCs molecules ) $U_0$ can take values ranging from  $U_0\approx0.01$ (at $g_{\rm cp}/aE_{\rm R,p}=1/2$, $r_d=200$) to $U_0\approx 5$ (at $g_{\rm cp}/aE_{\rm R,p}=5$, $r_d=1$), while still having a crystal and being in the tight binding limit. For a mixture with a mass ratio of $m_{\rm p}/m_{\rm c}\sim1/20$ the coupling $U_0$ is significantly larger, i.e. $U_0\approx 50$ (at $g_{\rm cp}/aE_{\rm R,p}=5$, $r_d=1$ for $m_{\rm p}/m_{\rm c}\sim1/20$).

\begin{figure}[t!]
\begin{flushright}
\includegraphics[width=.85\columnwidth]{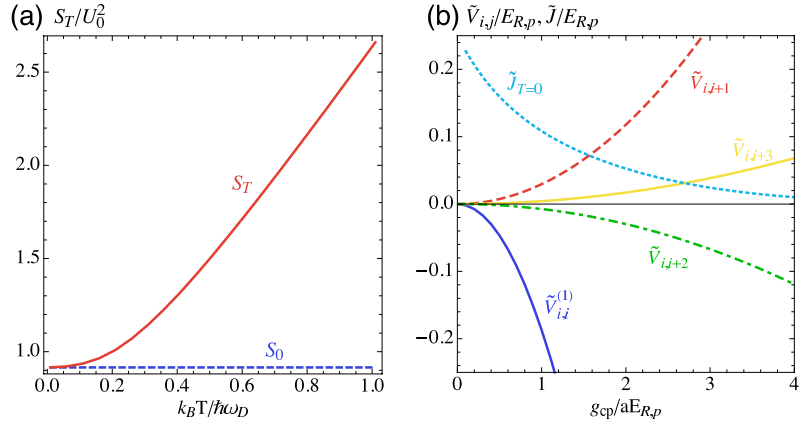}
\end{flushright}
\caption{\label{mod1fig1_100}
(a) The ratio $S_T/U_0^2$ as a function of the (dimensionless) temperature $k_{\rm B}T/\hbar\omega_{\rm D}$ with the dimensionless prefactor $U_0=(g_{\rm cp}/aE_{\rm R,p})/(m_{\rm p}/m_{\rm c})r_d^{3/4}$. The dashed line denotes the value $S_{T=0}/U_0^2$ and we find that for small temperatures $k_{\rm B}T/\hbar\omega_{\rm D}\ll 1$ the influence of the temperature on $S_T$ is negligible. $S_T$ depends quadratically on $U_0$ which realistically takes up values inbetween $10^{-2}$ and $10$. Thus we switch between the weak ($S_T\ll 1$) and strong ($S_T\gg 1$) coupling regimes by choice of the dimensionless prefactor $U_0$.
(b) The full particle-particle interaction $\tilde{V}_{ij}$ between two extra particles at interparticle distances $|i-j|=0,1,2,3$ for $r_d=100$ and $m_{\rm p}/m_{\rm c}=0.1$ as a function of the coupling $g_{\rm cp}$ in units of the particle recoil energy. Notice that the sign of the interaction alternates with every site and that the total interaction strength decreases  with the interparticle distance as $\propto 1/|i-j|^2$.}
\end{figure}

\begin{figure}[t!]
\begin{flushright}
\includegraphics[width=.85\columnwidth]{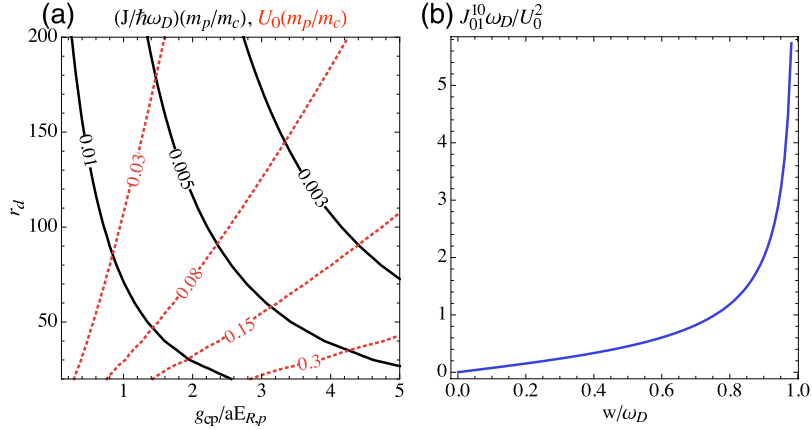}
\end{flushright}
\caption{\label{mod1fig1_5}
(a) Contour plots of the coupling ratio times the mass ratio $U_0\times(m_{\rm p}/m_{\rm c})$ (dashed contour lines) and of the ratio of the (bare) atomic tunneling rate and the Debye frequency of the crystal $J/\hbar\omega_{\rm  D}$ (solid contour lines) as a function of the atom-molecule coupling $g_{\rm cp}/aE_{\rm R,p}$ and the stiffness of the crystal $r_d$. The ratio $U_0(m_{\rm p}/m_{\rm c})$ determines the overall strength of the displacement amplitudes $u_q$ for a polaron, while the ratio $J/\hbar \omega_{\rm D}$ characterizes the serparation of crystal and interaction time (see text), and serves as ``the'' smallness parameter in the derivation of a master-equation \eref{masta01}. (b) The spectral density for the swapping of two particles on neighboring sites, $J_{01}^{10}(\omega)$, as a function of the frequency $\omega/\omega_{\rm D}$. It displays a Van Hove singularity at the Debye frequency $\omega_{\rm D}$ where it diverges as $\sim (\omega_{\rm D}-\omega)^{-1/2}$.}
\end{figure}

Due to the dragging of the surrounding phonon cloud, the hopping rate for the polarons, $\tilde J$, compared to the (bare) hopping rate, $J$, is suppressed by the factor $\tilde{J}/J=e^{-S_T}$, where the exponent is, cf.~\eref{st2},
\begin{equation}\label{STmod2}
\hspace{-.3cm}S_T=\frac{8\pi^4U_0^2}{(186\zeta(5))^{3/2}}\frac{a^2}{N}\sum_{q}\frac{q^2}{{\rm w}_{q}^3}\beta_{q}^2\sin^2\Big(\frac{qa}{2}\Big)\coth\Big(\frac{\hbar\omega_q}{2k_{\rm B}T}\Big).
\end{equation}
We notice that $S_T$ increases quadratically with $U_0$ (cf. the coupling $g_{\rm cp}$), whereas the ratio $S_{T}/U_0^2$ depends only on the temperature $T$ (in units of the Debye frequency), which is shown in \fref{mod1fig1_100}(a). In particular we remark that for $T=0$ the exponent scales with the coupling ratio $U_0$ as $S_{T=0}\approx0.9U_0^2$ [indicated by a horizontal dashed line in \fref{mod1fig1_100}(a)], while $S_T/U_0^2$ only weakly increases with the temperature $T$.\\

The phonon-coupling provides phonon mediated particle particle interactions of strength
\begin{equation}
\tilde{V}_{ij}^{(1)}=-\frac{4\pi^2}{93\zeta(5)}\frac{g_{\rm cp}^2}{E_{\rm R.p}}\frac{(m_{\rm c}/m_{\rm p})}{N r_d}\sum_{q}\frac{q^2}{{\rm w}_{q}^2}\beta_{q}^2\cos (qa|i-j|),
\end{equation}
for two polarons at sites $i$ and $j$, respectively. We notice that  $\tilde V_{ij}^{(1)}$ are temperature-independent and vary in sign and magnitude with the separation $i-j$, i.e. they are attractive for even $i-j$ and repulsive for odd $i-j$, while their absolute value decreases with increasing inter-polaron separation $|i-j|$. In \fref{mod1fig1_100}(b) we show the leading contributions for the total off-site shifts, which are purely induced by the phonons, $\tilde V_{ij}=\tilde V_{ij}^{(1)}$ for $i\neq j$, as a function of $g_{\rm cp}/aE_{R,p}$, cf. the leading off-site terms decay as $V_{i,j}/2E_{\rm p}\approx 0.16/|i-j|^2$. In addition we also plot the corresponding (modified) hopping rate $\tilde J$ for zero temperature $T=0$. We notice that near $g_{\rm cp}/aE_{R,p}\sim 1.6$ the nearest neighbor interactions become comparable with the effective hopping rate, $\tilde{V}_{i,i+1}\sim\tilde J$.

For bosonic particles, the phonon-mediated interactions also include an attractve on-site shift, the value of which is exactly twice the polaron shift, 
\[\tilde V_{ii}^{(1)}=-2E_{\rm p}\approx -1.54\times  U_0^2\hbar\omega_{\rm D}.\]
This can (in principle) lead to a collapse of the system, as it favors the piling up of polarons at a single site. However, since the full interactions comprise the bare {\em and} the phonon-mediated interaction, the total onsite shift, $\tilde V_{ii}=V_{ii}-2E_{\rm p}$, is positive for $V_{ii}/2>E_{\rm p}$, which we require to ensure the stability of the bosonic system. We remark that by resorting to tune $g_{\rm pp}$ via a Feshbach resonance, one can tune the onsite-shift up to $V_{ii}\sim\Delta$, without breaking the single-band approximation in the Hubbard model \eref{eq:1}. Thus we notice that in principle the stability of the system can be guaranteed, provided $E_{\rm p}<(V_{ii}/2)<\Delta/2$.

\subsubsection{Corrections to the extended Hubbard model}

In the following we are interested in higher-order corrections to the effective Hubbard model $H_{\rm S}$ of $\eref{heff0}$, which we derived in \Sref{section:2} in terms of the  spectral densities $J^{kl}_{ij}(\omega)$ for correlated nearest-neighbor hopping events $i\rightarrow j$ and $k\rightarrow l$. For our atomic-crystalline mixture we find from \eref{specda} that the latter are given by
\begin{equation}\label{eq:51}
J_{ij}^{kl}(w)\approx\frac{16\pi^3U_0^2}{(186\zeta(5))^{3/2}}\frac{(q_wa)^2}{(w/\omega_{\rm D})^3\sqrt{\omega_{\rm D}^2-w^2}}g_{ij}^{kl}(q_w),
\end{equation}
where we took $\beta_q\approx1$ and $\omega_q\approx \omega_{\rm D}\sin(qa/2)$ and $q_\omega\equiv {\rm arcsin}(\omega/\omega_{\rm D})/a$. The spectral density for the ``swapping'' of two particles on neighboring sites, $J^{10}_{01}(\omega)$ is shown in \Fref{mod1fig1}(c) and shows a Van Hove singularity at $\omega\rightarrow\omega_{\rm D}$, due to the $1/\sqrt{\omega_{\rm D}-\omega}$ divergence of the density of states for the crystal phonons.\\

\begin{figure}[t!]
\begin{flushright}
\includegraphics[width=.8\columnwidth]{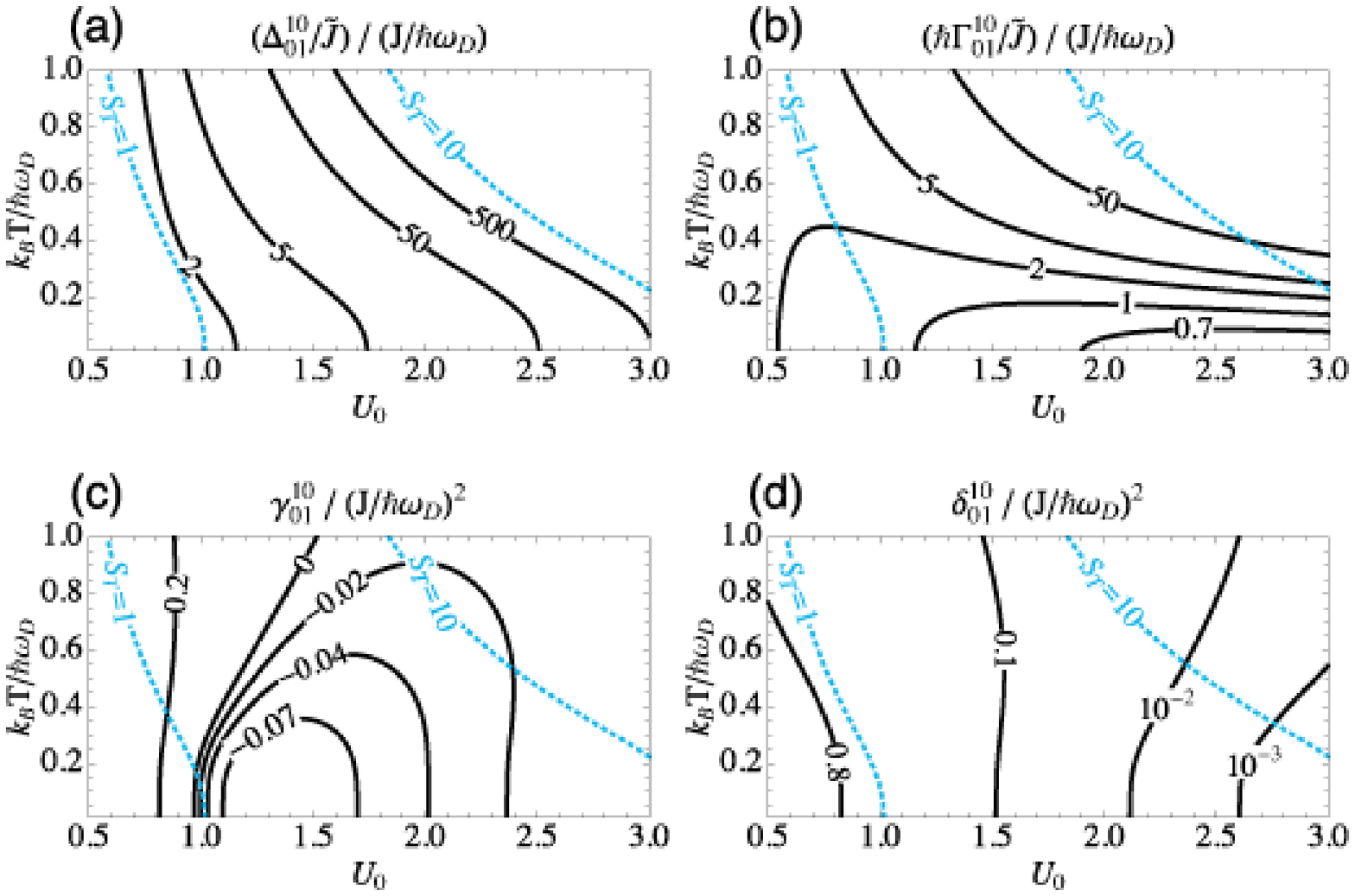}
\end{flushright}
\caption{\label{mod1fig3}
Leading corrections to the extended Hubbard model for a atomic-crystalline mixture in the strong coupling regime, $S_T\gg1$, as function of the temperature of the crystal $T$ and the coupling ratio $U_0$: The solid lines indicate contours for (a) the incoherent ``rate'' $\Gamma_{01}^{10}$ and (b) the coherent ``shift'' $\Delta_{01}^{10}$ (in units the effective hopping rate $\tilde J$), (c) the incoherent coefficient $\gamma_{01}^{10}$ and (d) the coherent coefficient $\delta_{01}^{10}$ as obtained from the strong-coupling approximation \eref{scDelta}-\eref{scdelta}, respectively. The two dashed lines in each panel represent the contours where $S_T=1$ and $S_T=10$, respectively. They designate the area, where $S_T>1$, and thus the strong coupling approximations holds. Notice that in (a,b) $\Gamma/\tilde J$, $\Delta/\tilde J$ are divided by the small ratio $J/\hbar\omega_{\rm D}$, while in (c,d) $\gamma$, $\delta/\tilde J$ are divided by the even smaller ratio $(J/\hbar\omega_{\rm D})^2$.}
\end{figure}

\begin{figure}[t!]
\begin{flushright}
\includegraphics[width=.85\columnwidth]{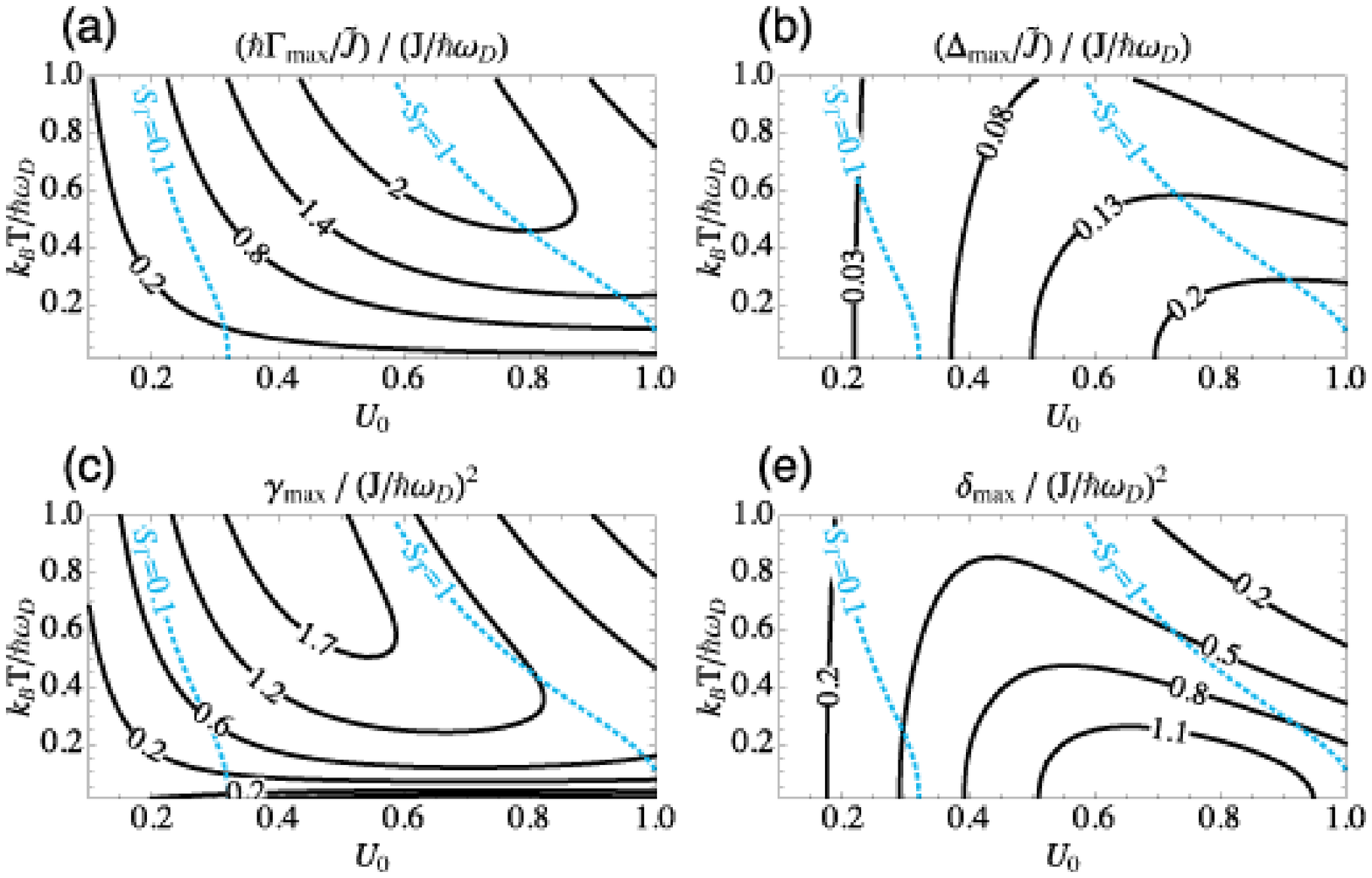}
\end{flushright}
\caption{\label{mod1fig4} Leading corrections to the extended Hubbard model for an atomic-crystalline mixture in the weak coupling regime, $S_T\ll1$, as function of the temperature of the crystal $T$ and the coupling ratio $U_0$: The solid lines indicate contours for the largest (a) incoherent rate $\Gamma_{\rm max}$ and (b) coherent shift $\Delta_{\rm max}$ (in units the effective hopping rate $\tilde J$), (c) incoherent coefficient $\gamma_{\rm max}$ and (d) coherent coefficient $\delta_{\rm max}$ as obtained from the weak-coupling approximation \eref{eq:Deltawc}-\eref{eq:deltawc}, respectively. The two dashed lines in each panel represent the contours where $S_T=0.1$  and $S_T=1$, designating the area, where $S_T<1$, and thus the weak coupling approximations hold. Notice that in (a,b) $\Gamma/\tilde J$, $\Delta/\tilde J$ are divided by the small ratio $J/\hbar\omega_{\rm D}$, while in (c,d) $\gamma$, $\delta/\tilde J$ are divided by the even smaller ratio $(J/\hbar\omega_{\rm D})^2$.}
\end{figure}

{\it Strong coupling limit $S_T \gg 1$:} The value $S_T$ is determined from equation~\eref{STmod2}. Values $S_T \gg 1$ are obtained for large coupling ratios $U_0$ and/or high temperatures $T$. However, already for a (reasonably small) ratio $U_0>1.1$ we have $S_T>1$ at $T=0$ and thus we are in the strong-coupling regime for all temperatures $T\geq0$. In this limit the main corrections to the Hubbard model~\eref{heff0} are due to "swap" processes. The corresponding rates and coefficients for $S_T\gg1$ are well approximated by the expressions \eref{scGamma}-\eref{scdelta}, which are shown in \fref{mod1fig3} and \fref{mod1fig4} as a function of the coupling ratio $U_0$ and the temperature $T$. Notice that in \eref{scDelta}-\eref{scdelta} the parameter $B=4\pi^6U_0^2/3(186\zeta(5))^{3/2}\approx 0.48U_0^2$ while the ratio $A_T/B$ exceeds $(k_{\rm B}T/\hbar\omega_{\rm D})\tanh(\hbar\omega_{\rm D}/2k_{\rm B}T)$.

In \fref{mod1fig3}(a-d) we show the leading contributions in the strong-coupling regime as a function of the ratio $U_0$ and the temperature $T$. In particular, panels (a), (b), (c), and (d) are contour plots of the incoherent rate $\hbar \Gamma_{01}^{10}/\tilde J$, the coherent shift $\Delta_{01}^{10}/\tilde J$, $\gamma$ and $\delta$, respectively, as obtained from the strong coupling approximation \eref{scDelta}-\eref{scdelta}. The two dashed lines in each panel signal where $S_T=10$ and $S_T=1$, and the strong-coupling approximation is valid ($S_T > 1$). We remark that in panels (a,b) $\Gamma/\tilde J$, $\Delta/\tilde J$ are divided by the (small) ratio $J/\hbar\omega_{\rm D}$, while in (c,d) $\gamma$ and $\delta$ are divided by the even smaller ratio $(J/\hbar\omega_{\rm D})^2$. We notice that in the spirit of the master equation approach, all quantities in the figures are plotted at finite temperature. The figure shows that, in a wide range of parameters, corrections to the coherent time evolution determined by $H_S$ can be made small in the strong coupling regime.\\

{\it Weak coupling limit $S_T \ll 1$:} In \Fref{mod1fig4} we show the leading corrections to the extended Hubbard model $H_S$ as a function of the coupling ratio $U_0$ and the temperature $T$, as obtained from the weak-coupling approximations \eref{eq:Deltawc}-\eref{eq:deltawc}. In particular, the solid lines now indicate contours of the largest value attained for (a) $\Gamma/\tilde J$, (b) $\Delta/\tilde J$, (c) $\gamma$ and (d) $\delta$. In panels (a,b), $\Gamma/\tilde J$, and $\Delta/\tilde J$ are divided by the small ratio $J/\hbar\omega_{\rm D}$, while $\gamma$ and $\delta$ in (c,d) are divided by the even smaller ratio $(J/\hbar\omega_{\rm D})^2$. The two dashed lines indicate $S_T=0.1$ and $S_T=1$, where the latter delimits the range of validity of the weak-coupling expressions \eref{eq:Deltawc}-\eref{eq:deltawc} (e.g., $U_0\lesssim 1$ for $T=0$ while $U_0\lesssim 1/2$ for $T=\hbar\omega_{\rm D}$). We notice that in the area where $S_T\lesssim 0.1$ the ratios shown in panels (a-d) are smaller than $\approx 1$, and thus all corrections are strongly suppressed compared to $\tilde J$, provided $J\ll\hbar\omega_{\rm }$.

\subsection{Polar molecules interacting with a one-dimensional crystal}
As a second configuration, we consider a setup where polar molecules of a second species are trapped at a distance $b$ from the crystal tube, under one-dimensional trapping conditions [see \fref{figs:fig1}(c)]. An external electric field aligns all dipoles in the direction perpendicular to the plane containing the two tubes. Molecules trapped in the two different tubes interact via long-range dipole-dipole interactions.

\subsubsection{Tight binding limit and Hubbard models}
\begin{figure}[t!]
\begin{flushright}
\includegraphics[width=.85\columnwidth]{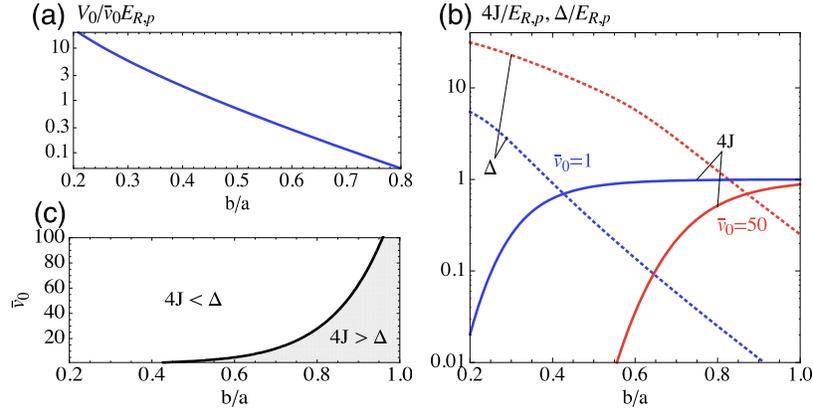}
\end{flushright}
\caption{\label{figMod10}
In panel (a) we plot the potential depth $V_0$ as a function of $b/a$ for different values of $r_d$ in units of the crystal recoil energy to give an idea about the extra-particle crystal interaction potential which determines the bandstructure. The hopping amplitude $4J$ and the gap to the first excited band $\Delta$ are shown in panel (b) as a function of $b/a$ in units of the particle recoil energy. The bandstructure strongly depends on the ratio of the particle-crystal interaction energy over the kinetic particle energy $\bar{v}_0 = (d_{\rm p}/d_{\rm c})r_d/(m_{\rm p}/m_{\rm c})$. The hopping amplitude decreases while the gap increases rapidly with increasing coupling strength and $\bar{v}_0$. The single band model is only valid where the gap exceeds the bandwidth $\Delta>4J$.}
\end{figure}

For crystal molecules fixed at the equilibrium positions with lattice spacing $a$, the particles (that is, the molecules of the second species) feel the following periodic potential
\begin{eqnarray}\label{vcp}
V_{\rm cp}(x)=\frac{d_{\rm c}d_{\rm p}}{a^3}\sum_j\frac{1}{[(b/a)^2 + (x/a-j-1/2)^2]^{3/2}},
\end{eqnarray}
where $d_{\rm p}$ is the induced dipole moment of the second-species molecules. The potential above has a depth
\begin{eqnarray}\label{vcp1}
V_{0}\equiv V_{{\rm cp}}(a/2)-V_{{\rm cp}}(0)\sim \bar{v}_0 e^{-3 b/a} E_{\rm R,p} /(b/a)^3,\nonumber
\end{eqnarray}
which determines the band-structure for the particles, with
\begin{equation}
\bar{v}_0=(d_{\rm p}/d_{\rm c})(m_{\rm c}/m_{\rm p})r_d, \label{eq:v0}
\end{equation}
and  $E_{\rm R,p}=\hbar^2\pi^2/2m_{\rm p}a^2$ the particle recoil energy. The lattice depth $V_{0}$ is shown in Fig.~\fref{figMod10}(a) to have a comb-like structure for $b/a<1/4$, since the particles resolve the
individual molecules forming the crystal, while it is sinusoidal for
$b/a\gtrsim1/4$. \Fref{figMod10}(b) shows the width $4J$ of the lowest-energy band, with $J/E_{\rm R,p}\sim (V_0/E_{\rm R,p})^{3/4}e^{-2
\sqrt{V_0/E_{\rm R,p}}}$ for $b/a\gtrsim1/4$, together with the energy gap $\Delta \simeq (4 V_0 E_{\rm
R,p})^{1/2}$, as a function of $b/a$ and for $\bar v_0=1$ and 50.
For a single particle, the single-band model is valid for $4J < \Delta$. \Fref{figMod10}(c) is a contour plot of the regimes of validity of the single-band model as a function of $b/a$ and $\bar v_0$.

When more particles are considered, the strong dipole-dipole repulsion between the particles acts as an effective hard-core constraint
\cite{Buechler07NPHYS}. We find that for $4J<\Delta$ and $d_{{\rm p}}\ll d_{{\rm c}}$ the bare off-site interactions satisfy
$V_{ij}\sim d_{{\rm p}}^{2}/(a|i-j|)^{3}<\Delta$, and thus the single-band model is still valid.\\

\begin{figure}[t!]
\begin{flushright}
\includegraphics[width=.85\columnwidth]{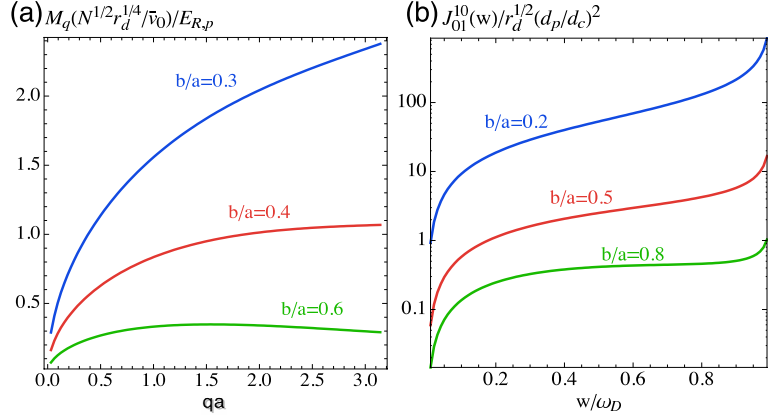}
\end{flushright}
\caption{\label{mod2fig2}
Figure (a) shows the dependence of particle-phonon coupling $M_q$ on the quasi momentum $qa$ for different values of $b/a$. Notice that as expected the coupling strength increases with small $b/a$ and large $\bar{v}_0$. For small $q$ it tends to zero like $q^{1/2}/(b/a)^2$. Figure (b) shows how the spectral density behaves as a function of $w$ for different values of $b/a$ on a logarithmic scale. It depends strongly on the ratio $b/a$ and tends to zero like $w$ for small frequncies.}
\end{figure}

In this configuration, the particle-phonon coupling as obtained from equation~\eref{phcoupl} is given by
\begin{equation}\label{mqmod1}
M_q=\frac{d_{\rm p}d_{\rm c}}{ab}\sqrt{\frac{2\hbar}{N m_{\rm c}\omega_q}}q^2\mathcal{K}_1(b|q|)\beta_q
\end{equation}
where $\mathcal{K}_1$ denotes the modified Bessel function of the second kind, and  $\beta_q=\int dx e^{iqx}|w_0(x)|^2$, with  $w_0(x)$ the lowest-band Wannier functions. \Fref{mod2fig2}(a) shows that
for $b/a$ small-enough, such that the single-band approximation is fulfilled for all $\bar v_0$ [see \fref{figMod10}(c) above], $M_q$ becomes peaked at large quasimomenta $q$.
We notice that consistency with the requirement of a stable crystal implies that the variance of the fluctuations of the crystal molecules around their equilibrium positions induced by the presence of a particle localized at a site $j$, $\langle\delta v_{ij}\rangle$, be small compared to the interparticle distance $a$, that is $\langle\delta v_{ij}\rangle/a < 1$. For a given ratio $ d_{\rm p}/d_{\rm c}$, this limits how small the ratio $b/a$ can realistically be, in order to avoid that the inter-species interactions
destroy the crystalline structure~\cite{PupilloPRL2008}. For example, for a ratio $d_{\rm p}/d_{\rm c}\approx 0.1$ the ratio $b/a$ can be as small as $b/a\approx 0.2$.

%
%

\subsubsection{Extended Hubbard model for molecular polarons inside a dipolar crystal}
\begin{figure}[t!]
\begin{flushright}
\includegraphics[width=.85\columnwidth]{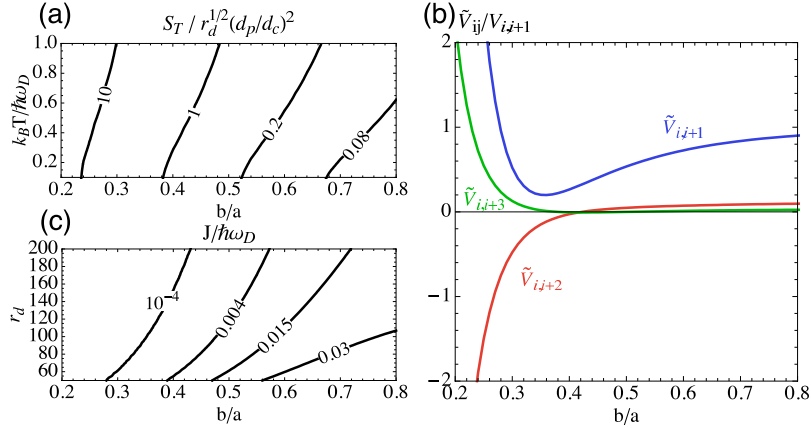}
\end{flushright}
\caption{\label{mod2fig3}
The quantity $S_T$ depends on the temperature $k_{\rm B}T$, the ratio $b/a$ and is proportional to $r_d^{1/2}(d_{\rm p}/d_{\rm c})^2$. In figure (a) we show how $S_T$ behaves with the dimensionless temperature $k_{\rm B}T/\hbar\omega_{\rm D}$ and the ratio $b/a$. By fixing the dipole ratio and $r_d$ we can determine where the weak and strong coupling regimes are valid. Figure (b) shows the full particle-particle interaction $\tilde{V}_{i,j}$, the sum of the bare dipole-dipole repulsion with the phonon-mediated interaction, in units of the bare nearest neighbour interaction $V_{i,i+1}$. The full interaction decays with increasing distance and shows an alternating sign for small ratios $b/a$ where the phonon-mediated interaction $V_{i,j}^{(1)}$ dominates. In figure (c) we show a contour of the separation of bath and interaction timescales, $J/\hbar\omega_{\rm D}$, that play an importent role when discussing the validity of our model and the amplitude of the perturbative corrections.}
\end{figure}

We continue by determining the modified Hubbard parameters $\tilde{J}$ and $\tilde{V}_{ij}$ for this configuration. Here, the parameter $S_T$, which determines the regime of interactions, is given by
\begin{equation}\label{stmod2}
S_T=\frac{32r_d^{1/2}(d_{\rm p}/d_{\rm c})^2}{(186\zeta(5))^{3/2}}\frac{1}{N}\sum_{q}\frac{(qa)^4\mathcal{K}_1^{2}(b|q|)}{{(b/a)^2\rm w}_q^3}\beta_q^2 \sin^2\Big(\frac{qa}{2}\Big)\coth\Big(\frac{\hbar\omega_q}{2k_{\rm B}T}\Big).
\end{equation}
The latter depends strongly on the ratio $b/a$ and is proportional to $r_d^{1/2}(d_{\rm p}/d_{\rm c})^2$. For a given $r_d$ and $d_{\rm p}/d_{\rm c}$ ratio, the regimes of weak and strong coupling, $S_T \ll 1$ and $S_T \gg 1$, respectively, can be directly determined from~\fref{mod2fig3}, which is a contour plot of $S_T$ as a function of the dimensionless temperature $k_{\rm B}T/\hbar\omega_{\rm D}$ and the ratio $b/a$. As in the previous model $S_T$ increases with increasing temperature and  particle-phonon coupling (that is, with decreasing ratio $b/a$).

The phonon mediated interaction as determined from equation~\eref{eq:8} is given by
\begin{equation}
\tilde{V}_{ij}^{(1)}=\frac{16r_d(d_{\rm p}/d_{\rm c})^2}{93\zeta(5)\pi^2}E_{\rm R,c}\frac{1}{N}\sum_q \frac{(qa)^4\mathcal{K}_1^{2}(b|q|)}{(b/a)^2{\rm w}_q^2}\beta_q^2.
\end{equation}
As in the previous model, the phonon mediated interactions show oscillations which for $b/a \lesssim 1/4$ decay slowly as $1/|i-j|^2$ and are thus long-ranged. Depending on their sign, they can enhance or reduce the bare dipole-dipole repulsions between two second-species molecules. The phonon-mediated interaction is strong and dominates the full particle-particle interaction $\tilde{V}_{ij}$ for small $b/a$. This is shown in \fref{mod2fig3}(b) which is a plot of $\tilde{V}_{ij}/V_{ij}$ as a function of $b/a$, where for $b/a\lesssim 0.4$ the value of $\tilde{V}_{ij}/V_{ij}$ can even change sign. With increasing intertube distance, the phonon-mediated term becomes small compared to the bare value $V_{ij}$.

\subsubsection{Corrections to the extended Hubbard model}
In the following we are interested in coherent and incoherent corrections to the time evolution determined by the effective Hubbard Hamiltonian $H_{\rm S}$ of~\eref{heff0}. These can be calculated in terms of the spectral density, which has the form
\begin{equation}
J_{ij}^{kl}(w) = \frac{64(d_{\rm p}/d_{\rm c})^2 r_d^{1/2}}{\pi(186\zeta(5))^{3/2}}\frac{\omega_{\rm D}^3}{w^3}\frac{(q_wa)^4\mathcal{K}_1(b q_w)^2\beta_{q_w}^2}{(b/a)^2\sqrt{\omega_{\rm D}^2-w^2}}g_{ij}^{kl}(q_w),
\end{equation}
where we have used $\omega_q\approx \omega_{\rm D}\sin(qa/2)$, and  $q_wa = 2\arcsin(w/\omega_{\rm D})$. The spectral density tends to zero linearly for small $w$ and shows an integrable (van Hove) singularity at $w=\omega_{\rm D}$. The spectral density $J_{01}^{10}(w)$ is plotted in~\fref{mod2fig2} for a few values of $b/a$.\\

General expressions for the corrections $\Delta_q(T), \Gamma_q(T),  \gamma_q(T)$ and $\delta_q(T)$ are given in~\eref{Delta}-\eref{delta}. For the configuration that we consider here, they depend on $b/a$, the temperature $k_{\rm B}T$ and the dimensionless parameter $(d_{\rm p}/d_{\rm c})^2 r_d^{1/2}$.  The quantities $\Gamma_q(T)$ and $\Delta_q(T)$ are proportional to the ratio $J/\hbar\omega_{\rm D}$, while $\delta_q(T)$ and $\gamma_q(T)$ are proportional to $(J/\hbar\omega_{\rm D})^2$. Thus, for later convenience, in \fref{mod2fig3}(c) we plot $J/\hbar\omega_{\rm D}$ as a function of $b/a$ and $r_d$ for a realistic choice of the dipole and mass ratios, $d_{\rm p}/d_{\rm c}=0.2$ and $m_{\rm p}/m_{\rm c} = 0.5$, respectively. The figure shows that for this choice of parameters the ratio $J/\hbar\omega_{\rm D}$ is (much) smaller than one for all plotted values of $b/a$ and $r_d$, and in particular it is e.g. of order $\sim 10^{-4}$ for reasonable values $b/a \approx 0.4$ and $r_d \approx 150$.\\

\begin{figure}[t!]
\begin{flushright}
\includegraphics[width=.85\columnwidth]{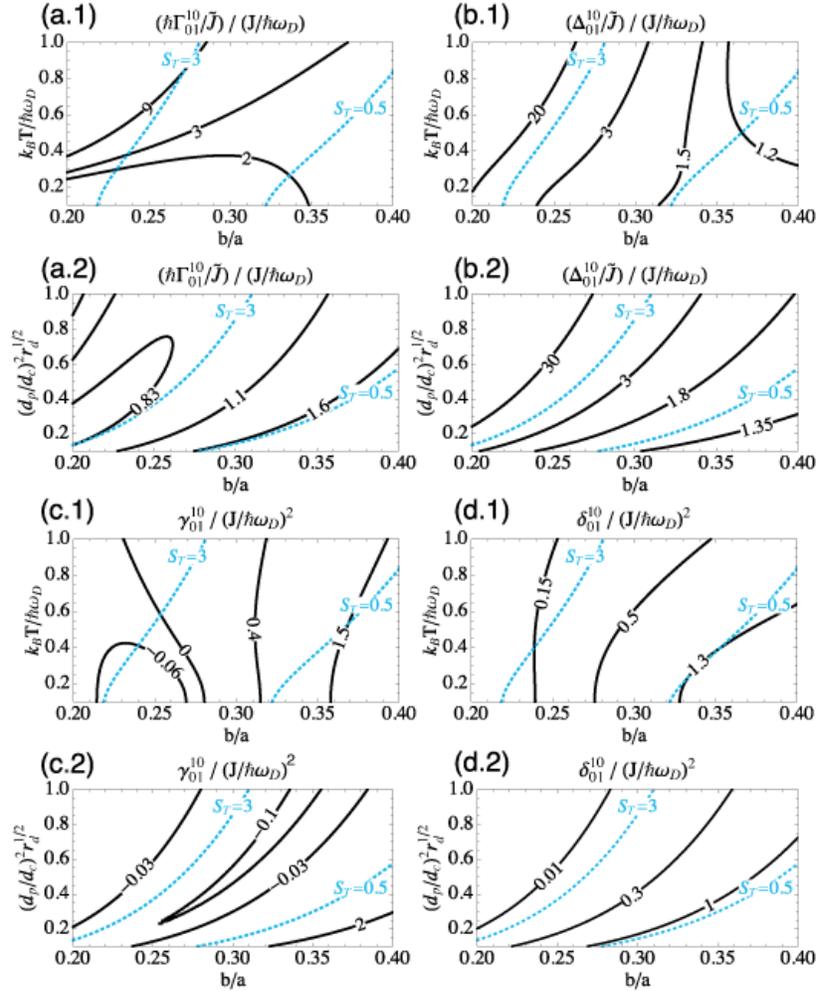}
\end{flushright}
\caption{\label{mod2fig4}
Ratios of the only non-negligible corrections that contribute to the eigenvalues as $\Gamma_q(T)\approx 2\Gamma_{01}^{10}(T)$, $\Delta_{q}(T)\approx 2\Delta_{01}^{10}(T)$, $\gamma_q(T) \ll 12 \gamma_{01}^{10}(T)$ and $\delta_q(T) \ll 12\delta_{01}^{10}(T)$ over the effective tunneling rate $\tilde{J}$ in the strong coupling limit for a single extra particle. These ratios are shown as functions of $b/a$, the dimensionless temperature $k_{\rm B}T/\hbar\omega_{\rm D}$ and $(d_{\rm p}/d_{\rm c})^2r_d^{1/2}$ in two distinct plots where we first fix $(d_{\rm p}/d_{\rm c})^2r_d^{1/2}=0.2$ and then the temperature as $k_{\rm B}T/\hbar\omega_{\rm D}=0.1$. The blue dotted lines show the value of $S_T$ and the strong coupling approximation breaks down below $S_T=0.5$. When including the separation of timescales $J/\hbar\omega_{\rm D}$ that is shown in \fref{mod2fig3}(c) we find that the corrections $\gamma_q(T)$ and $\delta_{q}(T)$ are negligible in the whole parameter regime. $\Gamma_q$ and $\Delta_q$ can however attain non negligible values for large $S_T$ and high temperatures.
}\end{figure}

{\it Strong coupling limit $S_T \gg 1$}: Here we are interested in giving examples of the importance of the corrections for realistic parameter regimes, compared to the characteristic energy $\tilde{J}$ of the polaronic Hamiltonian $H_{\rm S}$. Thus, in \fref{mod2fig4} we show contour plots of the quantities $\hbar\Gamma_{01}^{10}(T)/\tilde{J}$, $\Delta_{01}^{10}(T)/\tilde{J}$, $\gamma_{01}^{10}(T)$ and $\delta_{01}^{10}(T)$  as a function of $b/a$ and the dimensionless temperature $k_{\rm B}T/\hbar\omega_{\rm D}$ and the ratio  $(d_{\rm p}/d_{\rm c})^2 r_d^{1/2}$.

As explained in Section 2, these quantities correspond to the corrections for the "swap" process ($i=l$ and $k=j$), which is not suppressed exponentially by a factor $\propto\exp(-2 S_T)$, and is therefore the dominant correction in the strong-coupling limit $S_T \gg 1$~\cite{PupilloPRL2008,Alexandrov}. In particular, we can estimate $\Delta_{q}(T)\approx 2\Delta_{01}^{10}(T)$ and $\Gamma_q(T) \approx 2 \Gamma_{01}^{10}(T)$, while upper bounds for  $\delta_q(T)$ and $\gamma_q (T)$ can be estimated as  $\delta_{max}(T)<12\delta_{01}^{10}(T)$ and $\gamma_{max}(T)<12\gamma_{01}^{10}(T)$.

Panels (a1), (b1), (c1) and (d1) of \fref{mod2fig4} show results for  $\hbar\Gamma_{01}^{10}(T)/\tilde{J}$, $\Delta_{01}^{10}(T)/\tilde{J}$, $\gamma_{01}^{10}(T)$ and $\delta_{01}^{10}(T)$  as a function of $k_{\rm B}T/\hbar\omega_{\rm D}$, respectively, while the ratio $(d_{\rm p}/d_{\rm c})^2 r_d^{1/2}$ is fixed to the reasonable value 0.2. Panels (a2), (b2), (c2) and (d2) show the corrections as a function of  $(d_{\rm p}/d_{\rm c})^2 r_d^{1/2}$, with the temperature fixed to the value $k_{\rm B}T/\hbar\omega_{\rm D}=0.1$.

%

\begin{figure}[t!]
\begin{flushright}
\includegraphics[width=.85\columnwidth]{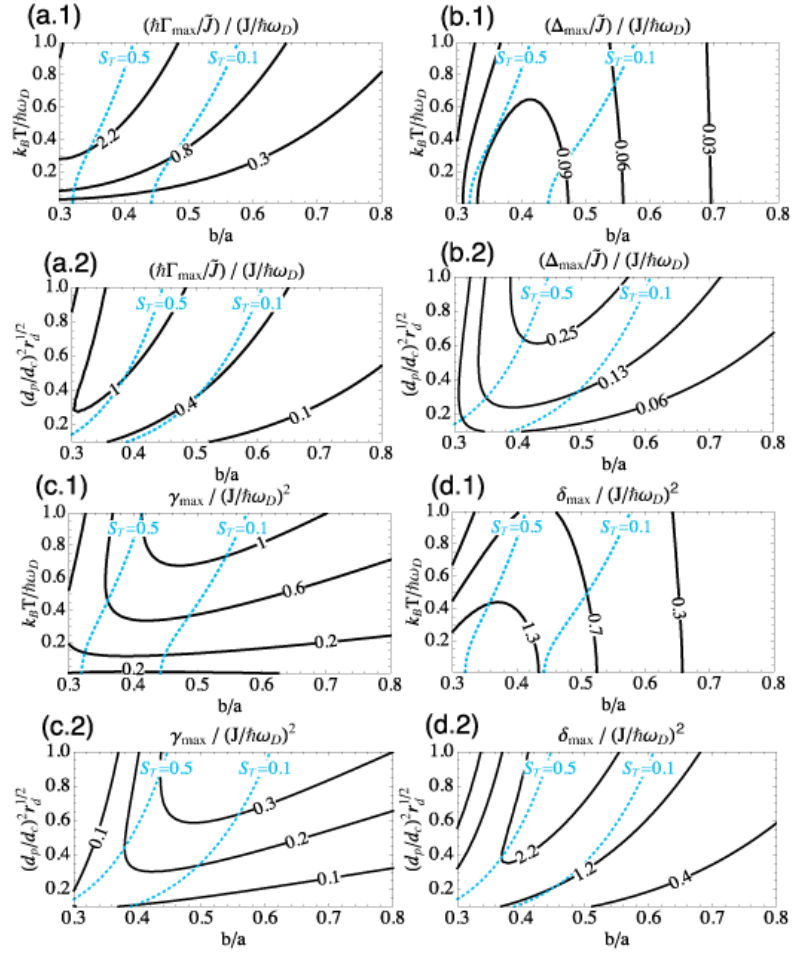}
\end{flushright}
\caption{\label{mod2fig5}
Ratios of the maximal eigenvalues of the corrections to the master equation over the effective tunneling rate $\tilde{J}$ in the weak coupling limit, $S_T\ll 1$, for a single extra particle. They plotted as functions of the ratio $b/a$, the dimensionless temperature $k_{\rm B}T/\hbar\omega_{\rm D}$ and $(d_{\rm p}/d_{\rm c})^2r_d^{1/2}$ in two distinct plots where we first fix $(d_{\rm p}/d_{\rm c})^2r_d^{1/2}=0.2$ and then the temperature as $k_{\rm B}T/\hbar\omega_{\rm D}=0.1$. The blue dotted lines outline the value of $S_T$ and the wek coupling approximation breakes down for values of $S_T > 0.5$. When taking the separation of timescales into account that is outlined in \fref{mod2fig3} we find that for $S_T< 0.5$ all the corrections to the master equation are negligiable compared to $\tilde{J}$.
}\end{figure}

\Fref{mod2fig4}(a.1)-(d.2) show that, for reasonably small $J/\hbar \omega_{\rm D}$ [see \fref{mod2fig3}(c)], $\gamma_{01}^{10}$ and $\delta_{01}^{10}$ tend to remain small in the range of shown parameters at finite $T$. However, the rate $\hbar\Gamma_{01}^{10}$ [panels (a.1)-(a.2)] and the energy $\Delta_{01}^{10}$ [panels (b.1)-(b.2)] can exceed the effective tunneling rate $\tilde J$ when the latter is strongly suppressed for strong couplings $S_T\gg 1$. While large $\hbar\Gamma_{01}^{10}$ can in principle lead to significant decoherence, and thus a transition from coherent hopping to thermally-activated hopping for large enough temperatures, we find that for reasonable temperatures $k_{\rm B}T/\hbar\omega_{\rm D} \lesssim 0.1$ these processes are strongly suppressed. On the other hand, the self-energies $\Delta_{01}^{10}(T)$ can lead to significant modifications to the coherent-time evolution determined by $H_{\rm S}$ by providing next-nearest neighbor hopping and sizeable off-site interactions in the strong coupling regime $S_T \gg 1$.\\

{\it Weak coupling limit $S_T \ll 1$}:
\Fref{mod2fig5} show the corrections to the coherent-time evolution given by $H_{\rm S}$ as a function of $b/a$ and the dimensionless temperature $k_B T/ \hbar \omega_{\rm D}$ [panels (a.1),(b.1),(c.1) and (d.1)] and the ratio  $(d_{\rm p}/d_{\rm c})^2r_d^{1/2}$ [panels (a.2),(b.2),(c.2) and (d.2)], in the regime of parameters where the single-band approximation is valid. We find that the weak coupling limit covers most of the accessible parameter regime for $b/a$. \Fref{mod2fig5} shows the maximal eigenvalues $\hbar \Gamma_{\max}$, $\Delta_{\max}$, $\gamma_{\max}$ and $\delta_{\max}$ for all four corrections. The central result here is that we find that the latter are negligible compared to the coherent hopping $\tilde{J}$ in the region where the weak coupling expansion is valid. In the figure, as a reference to identify how good the "weak-coupling" expansion is we also plot the values of $S_T$ for the various regimes of parameters.\\

Finally, we conclude this section by providing an example of the regime of validity of our model configuration. We consider a crystal of ${\rm Sr}{\rm O}$ where second-species molecules are ${\rm K}{\rm Rb}$. The dipole and mass ratios are $d_{\rm p}/d_{\rm c}\approx 0.08$ and $m_{\rm p}/m_{\rm c}\approx 1.2$, respectively. We find that our treatment properly accounts for the system dynamics for separations $0.2 \lesssim b/a \lesssim 0.7$, provided $r_d\gtrsim 80$.

\section{Conclusion}
In this work we studied the realization of lattice models in mixtures of cold atoms and polar  molecules, where a first molecular species is in a crystalline configuration and provides a periodic trapping potential for the second (atomic or molecular) species. We have treated the system dynamics in a master equation formalism in the Brownian motion limit for slow, massive, particles embedded in the molecular crystal with fast phonons. In a wide regime of parameters the reduced system dynamics corresponds to coherent evolution for particles dressed by lattice phonons, which is well described by extended Hubbard models. For two realistic one-dimensional setups with atoms and molecules these lattice models display phonon-mediated interactions which are strong and long-ranged (decaying as $1/|i-j|^2$). The sign of interactions can vary with distance from repulsive to attractive, which can possibly lead to the realization of interesting phases of interacting polarons in one dimensional dipolar crystals. This study, and extensions to two dimensions will be the subject of future work.

\section{Acknowledgments}
We acknowledge funding from the European Union through the STREP FP7-ICT-2007-C project NAME-QUAM (Nanodesigning of  Atomic and MolEcular QUAntum Matter) and the Austrian Science Foundation (FWF).

\appendix
\section{Hamiltonian for particles moving in a crystal}\label{AppA}
It is the aim of this section to derive the Hamiltonian~\eref{eq:1} starting from a generic mixture of two
interacting species of atoms or molecules.
\subsection{Effective continuum Hamiltonian}
A mixture of two interacting species is confined to one or two dimensions by a strong optical trapping
potential. In \cite{BuechlerPRL2007} it is shown how such a trapped system can be reduced to an effective
lower dimensional model in the low energy limit. The effective one- or two-dimensional Hamiltonian reads
\begin{equation}\label{eq:h0}
H_{\rm eff} = H_{\rm c} + H_{\rm p} + H_{\rm cp}.
\end{equation}
Here $H_{\rm c}$ describes the effective motion of the crystal particles, $H_{\rm p}$ the dynamics of the
extra particles with an extra particle- crystal particle interaction given by $H_{\rm cp}$. From Hamiltonian
\eref{eq:h0} we identify
\begin{eqnarray}
H_{\rm c}&=&\sum_i\frac{{\bf P}^2_i}{2m_{\rm c}}+\frac{1}{2}\sum_{i\neq j}V_{\rm cc}({\bf R}_i-{\bf
R}_j),\label{hc}\\ H_{\rm p}&=&\sum_i\frac{{\bf p}^2_i}{2m_{\rm p}}+\frac{1}{2}\sum_{i\neq j}V_{\rm pp}({\bf
r}_i-{\bf r}_j),\label{hp}\\ H_{\rm cp}&=&\sum_{i,j}V_{\rm cp}({\bf r}_i-{\bf R}_j).\label{hcp}
\end{eqnarray}
where we denote ${\bf p}_i$(${\bf r}_i$) and ${\bf P}_i$(${\bf R}_i$) the momentum (position) of extra
particles and crystal particles with masses $m_{\rm p}$ and $m_{\rm c}$,  respectively. The sums range over
all the respective particles in the mixture. The interaction between two particles from the species
$\alpha,\beta \in \{{\rm c},{\rm p}\}$ in a distance ${\bf r}$ from each other is denoted by
$V_{\alpha\beta}({\bf r})$.

We take the continuum Hamiltonian \eref{eq:h0} as the starting point for the following discussion and derive
our lattice model from it.
\subsection{Crystal Hamiltonian}\label{settingup1}
In the crystalline phase the particles are characterized by small fluctuations of their positions ${\bf R}_j$
around their equilibrium positions ${\bf R}_j^{0}$. Thus we expand the potential in a Taylor series to second
order in the displacements ${\bf u}_j={\bf R}_j-{\bf R}_j^0$ about the equilibrium positions \cite{Mahan} as
\begin{equation}
V_{\rm cc}({\bf R}_i-{\bf R}_j)\approx V_{\rm cc}({\bf R}_i^0-{\bf R}_j^0) + {\bf u}_i{\sf
D}_{ij}{\bf u}_j,
\end{equation}
with the tensor ${\sf D}_{ij}$
\begin{equation}\label{eq:DTensor}
{\sf D}_{ij}\equiv\frac{1}{2}\nabla\otimes\nabla V_{\rm cc}({\bf R}_i^0-{\bf R}^0_j),
\end{equation}
which is readily diagonalized and provides the dispersion relation $\omega_{{\bf q},\lambda}$ for the phonons
with quasimomentum ${\bf q}$ and polarization ${\bf e}_\lambda$. We express the displacement ${\bf u}_i$ in
terms of the bosonic creation (annihilation) operators $a_{{\bf q},\lambda}^{\dag}$ ($a_{{\bf q},\lambda}$)
for the corresponding phonons as
\begin{equation}
{\bf u}_i =\sum_{{\bf q},\lambda} \sqrt{\frac{\hbar}{2N m_{\rm c}\omega_{{\bf q},\lambda} }} e^{i {\bf q}{\bf
R}_i^{0}} {\bf e}_\lambda (a_{{\bf q},\lambda}+a_{-{\bf q},\lambda}^{\dag})\label{displacement},
\end{equation}
where $N$ denotes the total number of crystal particles. We write the Hamiltonian as
\begin{equation}\label{pshc}
H_{\rm c}=\sum_{{\bf q},\lambda}\hbar\omega_{{\bf q},\lambda}a_{{\bf q},\lambda}^{\dag}a_{{\bf q},\lambda}
\end{equation}
and can identify $m_{\rm c}\omega_{{\bf q},\lambda}^2$ from the eigenvalues of $D_{ij}$ from which we are able
to read off the phononic dispersion relation, see \ref{AppB}.
\subsection{Extra particles inside the crystal}
The dynamics of extra particles inside the crystal and their interaction with the crystal is described by
$H_{\rm p} + H_{\rm cp}$.

In analogy to the proceedings of the last section we are interested in the dynamics of a stiff crystal and
therefore we expand the particle-crystal potential in the small displacements ${\bf u}_i$ of crystal
constituents about their equilibrium positions, as

\begin{equation}
\fl\qquad V_{\rm p}({\bf r})\equiv\sum_iV_{\rm cp}({\bf r}-{\bf R}_i)\approx \sum_iV_{\rm cp}({\bf r}-{\bf R}_i^0)+
\sum_i{\bf u}_i{\bf \nabla}V_{\rm cp}({\bf r}-{\bf R}_i^0).\label{vex}
\end{equation}
Here we have introduced the potential $V_{\rm p}({\bf r})$ that an extra particle at position ${\bf r}$ feels
from the entire crystal. The lowest order provides a static periodic potential
 \[V_{\rm p}^{(0)}({\bf r}) = \sum_iV_{\rm cp}({\bf r}-{\bf R}_i^0).\]
We note that in writing equation~\eref{vex} we only retain the first non-vanishing correction to the static
trapping potential, i.e. the term linear in the displacement ${\bf u}_i$. Thereby we neglect terms of second
(and higher) order in ${\bf u}_j$, which are expected to (merely) provide a renormalization of the
phonon-spectrum, i.e. when including those terms in equation~\eref{eq:DTensor}. 

The extra particles that we consider are confined to a plane/tube parallel to a crystal plane/tube, as
pictured in \fref{figs:fig1}. Hamiltonian (\ref{hp}) together with the zeroth order contribution to the
particle-crystal interaction,
\begin{equation}\label{hpprime}
H_{\rm p}'=H_{\rm p} + \sum_iV_{\rm p}^{(0)}({\bf r}_i),
\end{equation}
describes interacting particles for which the entire crystal provides a static periodic trapping potential,
the ingredients for a simple Hubbard model. The particle crystal interaction then determines the band
structure.

We consider a single band model where extra particles cannot be excited to the second band that is separated
from the lowest band by an energy gap $\Delta$. Then in the low energy limit the extra particles localize at
single sites of position ${\bf r}_i^0$ and their wave functions become Wannier functions of the lowest band
$w_0({\bf r}-{\bf r}_i)$. In second quantization the field operator can then be expanded as
\[\psi_{\rm p}({\bf r}) = \sum_{i} c_i w_0({\bf r}-{\bf r}_i^0),\]
where $c_i^{\dag}$ and $c_i$ denote the creation and annihilation operators of extra particles at site $i$,
which obey the canonical bosonic (fermionic) commutation (anticommutation) relations for bosons (fermions).

For such a model, see \cite{JakschBH}, the dynamics of the extra particles is described by the hopping
amplitude $J$ that calculates from the overlap of the wavefunctions at two neighbouring sites and the
interaction between two particles located at sites $i$ and $j$ is given by
\[
V_{ij} \approx \int d{\bf r}d{\bf r}' |w_0({\bf r}-{\bf r}_i^0)|^2V_{\rm pp}({\bf r}-{\bf r}') |w_0({\bf
r}'-{\bf r}_j^0)|^2.
\]
Hamiltonian~\eref{hpprime} can then be written as
\begin{equation}
H_p' = -J\sum_{\langle i,j\rangle}c_i^{\dag}c_j +
\frac{1}{2}\sum_{i,j}V_{ij}c_i^{\dag}c_j^{\dag}c_jc_i\label{Hp}.
\end{equation}
The single band approximation is valid if all particle energies are (much) smaller than the gap, cf. $J,
V_{ij}\ll \Delta$. We remark, that the Debye frequency $\hbar\omega_{\rm D}$ in our models is typically
(much)larger than the gap while the particle phonon coupling is dominated at high frequencies, see discussion
below. In order to avoid excitations beyond the gap by the coupling we put a constraint on the temperature in
a way that all phonon modes with energies larger than $\Delta$, are essentially unoccupied. 

Let us now focus on the higher order terms of the interaction in equation~\eref{vex} which correspond to the
backaction of the crystal on the extra particles. The remaining part of the Hamiltonian is given by
\[H_{\rm cp}' = H_{\rm cp}-\sum_iV_{\rm p}^{(0)}({\bf r}_i)\]
and describes a dynamic coupling of the particles to the vibrations of the crystal. In second quantization
for the extra particles the remaining Hamiltonian is obtained (cf. \ref{appCPinteraction}) as
\begin{equation}
H_{\rm cp}'\approx \sum_{{\bf q},\lambda}M_{ {\bf q},\lambda}(a_{ {\bf q},\lambda}+a_{-{\bf
q},\lambda}^{\dag})\sum_j e^{i{\bf q}{\bf R}_j^0}c_j^{\dag}c_j,
\end{equation}
were we have introduced the particle-phonon coupling $M_{{\bf q},\lambda}$ given by
\begin{equation}\label{phcoupl}
M_{ {\bf q},\lambda}=\frac{({\bf q}{\bf e}_{\lambda})\beta_{\bf q}}{\sqrt{2Nm_c\omega_{{\bf
q},\lambda}/\hbar}}\sum_ie^{i {\bf q}{\bf R}_i^0}V_{\rm cp}({\bf R}_i^0).
\end{equation}
Here $\beta_{\bf q}$ denotes the Fourier transform of the modulus square of the Wannier function
\[
\quad\beta_{\bf q}=\int d{\bf r}|w_0({\bf r})|^2e^{i{\bf q}{\bf r}}.
\]
The function $\beta_{\bf q}$ accounts for the localization of the particles with a finite width in the static
trapping potential and thus, for small $q$, approaches $1$. Similarly, for the class of potentials we
consider (and discuss) in Section~\ref{rdc}, the Fourier transform of the particle-crystal potentials
approaches a finite value at ${\bf q}=0$. Therefore the particle-phonon coupling behaves like $q^{1/2}$ for
small momenta, while it is large for $q\sim\pi/a$.

The Hamiltonian of the entire model is given by $H_{\rm eff} = H_{\rm c} + H_{\rm p}' + H_{\rm cp}'$ and
reads
\begin{eqnarray}
H_{\rm eff} &=&-J\sum_{\langle i,j\rangle}c_i^{\dag}c_j
+\frac{1}{2}\sum_{i,j}V_{ij}c_i^{\dag}c_j^{\dag}c_jc_i \nonumber\\ &+&\sum_{{\bf q},\lambda,j}M_{{\bf
q},\lambda}e^{i{\bf q}{\bf R}_j^0}c_j^{\dag}c_j(a_{{\bf q},\lambda}+a_{-{\bf q},\lambda}^{\dag}) +\sum_{{\bf
q},\lambda}\hbar\omega_{{\bf q},\lambda}a_{{\bf q},\lambda}^{\dag}a_{{\bf q},\lambda}.\label{eq:10}
\end{eqnarray}
The crystal motion given by Hamiltonian~\eref{eq:10} corresponds to a set of uncoupled harmonic oscillators
under the influence of an extra particle density dependent force. Similarly to the problem of a charged
harmonic oscillator in a constant electric field \cite{Mahan} such a force displaces the crystal molecules
from their original position proportional to its strength while keeping the oscillation frequency fixed. A
crystal molecule at position ${\bf R}_i$ is therefore displaced by
\begin{equation}\label{maxDisplM}
{\bf v}_i=\sum_{ {\bf q},\lambda}\sqrt{\frac{2\hbar}{N m_{\rm c}\omega_{ {\bf q},\lambda}}}\frac{M_{{\bf
q},\lambda}}{\hbar\omega_{ {\bf q},\lambda}}e^{i{\bf R}_i^0}\eta_{\bf q},
\end{equation}
with the Fourier transform of extra particle density $\eta_{\bf q}=\sum_je^{-i {\bf q}{\bf
r}_j^0}c_j^{\dag}c_j$.

If the relative displacement between two crystal molecules $\delta{\bf v}_{ij}={\bf v}_i-{\bf v}_j$ at
neighbouring sites $i,j$ is small on the scale of the lattice constant $\langle \delta{\bf v}_{ij}\rangle\ll
a$ this effect can be neglected.
\section{Phonon spectrum and particle-phonon coupling}\label{AppB}
In this section we derive the specific form of the dispersion relation for one- and two-dimensional dipolar
crystals as well as the crystal-particle interaction.
\subsection{The phonon spectrum}
The crystalline phase is characterized by small displacements of the molecules from their equilibrium
positions ${\bf R}_i={\bf R}_i^0+{\bf u}_i$. A dipolar crystal that is trapped in one or two dimensions by an
optical trap with a trapping frequency $\omega_\perp$ is described by the following Hamiltonian
\begin{equation}\label{ilkj}
H_{\rm c} =\sum_i \frac{{\bf P}_i^2}{2m_{\rm c}} +\sum_{i,\mu}\frac{m_{\rm c}}{2}\omega_\perp^2{\bf
u}_{i,\mu}^2+\sum_{i\neq j}\frac{d_{\rm c}^2}{|{\bf R}_i-{\bf R}_j|^3}
\end{equation}
where ${\bf P}_i$ denote the momenta of the molecules. The dispersion relation is found from the second order
correction of the expansion of  the molecule-molecule interaction potential
\begin{eqnarray}
V_{\rm cc}^{(2)}({\rm R}_i-{\rm R}_j)&=D_{ij}({\bf u}_i-{\bf u}_j)^2\\ \mathwith
D_{ij}&=\frac{1}{2}\nabla\otimes\nabla V_{\rm cc}({\bf R}_i^0-{\bf R}_j^0).
\end{eqnarray}
We make an ansatz for the displacement as
\begin{equation}
{\bf u}_i =\sum_{{\bf q},\lambda} \sqrt{\frac{\hbar}{2N m_{\rm c}\omega_{{\bf q},\lambda} }} e^{i {\bf q}{\bf
R}_i^{0}} {\bf e}_\lambda (a_{{\bf q},\lambda}+a_{-{\bf q},\lambda}^{\dag}),
\end{equation}
and can identify $a_{{\bf q},\lambda}^\dag$ and $a_{{\bf q},\lambda}$ as creation and annihilation operators
of phonons with the polarization ${\bf e}_\lambda$ and quasi momentum ${\bf q}$ iff the matrix $D_{ij}$ is
diagonal. Then the polarization vectors for phonons with a dispersion $\omega_{{\bf q},\lambda}$ are the
eigenvectors of $D_{ij}$.

With the discrete Fourier transform
\begin{equation}
{\bf u}_j=\sqrt{\frac{1}{N}}\sum_{\bf q} e^{i {\bf q} {\bf R}_j^0}{\bf u}_{\bf q}\mathand{\bf
P}_j=\sqrt{\frac{1}{N}}\sum_{\bf q} e^{i {\bf q} {\bf R}_j^0}{\bf P}_{\bf q},
\end{equation}
we can write Hamiltonian~\eref{ilkj} as
\begin{eqnarray}
H_{\rm c}&=&\frac{1}{2m_{\rm c}}\sum_{\bf q}({\bf P}_{\bf q}{\bf P}_{-{\bf q}}+m_{\rm c}^2\omega_{\bf
q}^2{\bf u}_{\bf q}{\bf u}_{-{\bf q}}) =\sum_{{\bf q},\lambda}\hbar\omega_{{\bf q},\lambda}\left(a_{{\bf q},\lambda}^\dag a_{{\bf
q},\lambda}+\frac{1}{2}\right)\nonumber
\end{eqnarray}
and find the dispersion relation from calculating the eigenvalues of $D_{ij}$. The $+1/2$ contribution in the last 
equation is the vacuum zero point energy, a constant energy shift that is ommitted henceforth.

\subsubsection{Phonon spectrum in 1D:}\label{1Dphononspectrum}
In a 1D crystal tube the second order of the expansion in the displacement of the interaction potential is
given by
\begin{equation}\label{mimimi}
V_{\rm cc}^{(2)}({\bf R}_i-{\bf R}_j)=d_{\rm c}^2 \left(\begin{array}{ccc} 12&0&0\\ 0&-3&0\\ 0&0&-3\\
\end{array}\right)\frac{({\bf u}_i-{\bf u}_j)^2}{|{\bf R}_i^0-{\bf R}_j^0|^5}.\nonumber
\end{equation}
In $q$-space we find
\begin{eqnarray}
&\sum_{i,j}\frac{({\bf u}_i-{\bf u}_j)^2}{|{\bf R}_i^0-{\bf R}_j^0|^5}=\sum_{q}\frac{f_q}{a^5} {\bf u}_q{\bf u}_{-q},\\ \fl\qquad\quad\mbox{with}
 &f_q=\sum_{j>0} 4\sin^2(qaj/2)/j^5=2\zeta(5)-Li_5(e^{iqa})-Li_5(e^{-iqa}),\label{mhgd}
\end{eqnarray}
with $\zeta(5)$ the zeta function at $5$  and $Li_5(x)$ the polylogarithm of fifth order at $x$. Here we have used the
fact that in the crystal tube the equilibrium positions are given by ${\bf R}_i^0=a i$ with the lattice
spacing $a$.

Since $D_{ij}$ is diagonal in the canonical basis the polarization vectors are given by the canonical basis
vectors and when including the optical trapping potential, the second term on the right handside of
equation~\eref{ilkj}, the dispersion relations are found as
\begin{eqnarray}
\omega_{q,\parallel}&=&\sqrt{\frac{12d_{\rm c}^2 f_q}{a^5m_{\rm c}}},\label{dispersx}\\
\omega_{q,\perp}&=&\sqrt{\omega_\perp^2-\frac{3d_{\rm c}^2 f_q}{a^5m_{\rm c}}}.\label{dispersy}
\end{eqnarray}
The longitudinal dispersion relation $\omega_{q,\parallel}$ is acoustic while the two transversal ones
denoted by $\omega_{q,\perp}$ show an optical behaviour. The fact that the transversal dispersion relations
can become imaginary points towards an instability of the crystal, reflected by the requirement for a strong
transversal trapping, see Section~\ref{settingup}, Equations \eref{dispersx} and \eref{dispersy} show that
the optical (transversal) modes decouple from the longitudinal one if
\[\omega_\perp > \sqrt{\frac{15 d_{\rm c}^2}{a^5m_{\rm c}} }\]
is fulfilled.
\subsubsection{Phonon spectrum in 2D:}\label{2Dphononspectrum}
We consider a dipolar crystal that forms in the $xy$-plane. The expansion of the molecule molecule potential
to second order in the displacement gives
\begin{equation*}
\fl V_{\rm cc}^{(2)}(\Delta{\bf R}_{ij}) = \frac{3d_{\rm c}^2}{2}\frac{({\bf u}_i-{\bf u}_j)^2}{|\Delta {\bf
R}_{ij}^0|^7}\left(\begin{array}{ccc}5\Delta {R_{ij,x}^0}^2-|\Delta {\bf R}_{ij}^0|^2&5\Delta
R_{ij,x}^0\Delta R_{ij,y}^0&0\\5\Delta R_{ij,x}^0\Delta R_{ij,y}^0&5\Delta {R_{ij,y}^0}^2 -|\Delta {\bf
R}_{ij}^0|^2&0\\0&0&|\Delta {\bf R}_{ij}^0|^2\end{array}\right)
\end{equation*}
where we have used $\Delta {\bf R}_{ij}={\bf R}_i-{\bf R}_j, \Delta {\bf R}_{ij}^0={\bf R}_i^0-{\bf R}_j^0$
and similarly $\Delta R_{ij,x}= R_{i,x}^0-R_{j,x}^0, \Delta R_{ij,y}= R_{i,y}^0-R_{j,y}^0$.
introduce $\iota = i-j$ and can write
It is difficult to 
diagonalize the matrix in the last equation. We can however
use the block diagonal form of the matrix in the last equation and, including the optical trapping potential
as we have done in the 1D case above, write the dispersion relations as
\begin{eqnarray}
\omega_{{\bf q},\pm}&=&\sqrt{\frac{d_{\rm c}^2}{a^5 m_{\rm c}} f^\pm_{\bf q}},\\
\omega_{{\bf q},z}&=&\sqrt{\omega_\perp^2-\frac{d_{\rm c}^2}{a^5 m_{\rm c}}f_{\bf q}},
\end{eqnarray}
where $f^\pm_q$ denote the two eigenvalues of the $xy$-block of the matrix and
$f_q$ is given by equation~\eref{mhgd}. The eigenvalues $f^\pm_q$ give a longitudinal and a
transversal acoustic dispersion relation. As noted above we cannot write them in a closed form but an
approximate result may be obtained by including only particles in the interaction that are within some finite
range of each other. We show a numeric evaluation of the dispersion relation in \fref{figdispersion}.


\subsection{The particle-phonon coupling}\label{appCPinteraction}
We denote the crystal-particle interaction potential by $V_{\rm cp}({\bf r}-{\bf R}_j)$ where ${\bf r}$ and
${\bf R}_j$ denote the position of an extra particle and a crystal particle respectively. The full potential
is expanded up to first order in the displacement ${\bf R}_j={\bf R}_j^0+{\bf u}_j$ as
\begin{eqnarray}
V_{\rm cp}({\bf r})&=&\sum_{j}{\bf u}_j \boldsymbol{\nabla}V_{\rm cp}({\bf r}-{\bf R}_j^0)\\
&=&\frac{1}{(2\pi)^d}\sum_{{\bf q},{\bf k},j} \tilde{{\bf u}}_{{\bf k}}e^{i({\bf k}-{\bf q}){\bf r}_j^0}{\bf
q} e^{i{\bf q}{\bf r}}\tilde{V}_{\rm cp}({\bf q})
\end{eqnarray}
where $d$ denotes the dimension of the setup and we have used the discrete Fourier transform of the
interaction potential $V_{\rm cp}({\bf R}_i^0)=\sum_{{\bf q}}e^{i{\bf q}{\bf
R}_i^0}\tilde{V}_{\rm cp}({\bf q})/\sqrt{2\pi}^d$ and the displacement ${\bf u}_j = \sum_{{\bf
k}}e^{i{\bf k}{\bf R}_i^0}\tilde{u}_{{\bf k}}/\sqrt{2\pi}^d$. The Hamiltonian is found by integration of the
extra particle density $\rho({\bf r})$ over the interaction $H_{\rm I}=\int d{\bf r} \rho({\bf r})V_{\rm cp}({\bf
r})$ which gives
\begin{eqnarray}
H_{\rm I}=\sqrt{2\pi}^d\sum_{{\bf q}}\tilde{\rho}({\bf q})\tilde{V}_{\rm cp}({\bf q}){\bf q}\tilde{{\bf u}}_{{\bf
q}}.
\end{eqnarray}
The extra particle density $\rho({\bf r})$ is defined through the single particle operator, which reads in
site representation $\psi({\bf r})=\sum_mc_m\phi_m({\bf r})$ with $\phi_m({\bf r})$ the wavefunction of an
extra particle at site $m$, and the annihilation operator $c_m$ as $\rho({\bf r})=\psi^{\dag}({\bf
r})\psi({\bf r})$. In the tight binding limit the particles are strongly localized at sites and the overlap of the wavefunctions of particles at different sites is neglected. This approximation gives
\begin{eqnarray}
\tilde{\rho}({\bf q})&\approx&\sum_mc_m^{\dag}c_m\int\frac{d{\bf r}}{\sqrt{2\pi}^d}|\phi_m({\bf
r})|^2e^{i{\bf q}{\bf r}}\\ &=&\sum_mc_m^{\dag}c_m\int\frac{d{\bf r}}{\sqrt{2\pi}^d}|\phi_0({\bf
r})|^2e^{i{\bf q}({\bf r}+{\bf r}_m^0)}\\ &=&\frac{\beta_{{\bf q}}}{\sqrt{2\pi}^d}\sum_mc_m^{\dag}c_me^{i
{\bf q}{\bf r}_m^0}.
\end{eqnarray}
We can work out $\beta_{{\bf q}}=\int d{\bf r}|\phi({\bf r})|^2e^{i{\bf q}{\bf r}}$ by making a separation
ansatz for the wavefunction $\phi({\bf r})=\phi_x(x)\phi_y(y)\phi_z(z)$ and thus $\beta_{{\bf
q}}=\beta_{q_x}\beta_{q_y}\beta_{q_z}$. In the directions of confinement the wavefunction is taken to be
Gaussian $\phi_z(z)=e^{-z^2/2a_\perp^2}/\pi^{1/4}a_\perp^{1/2}$ with
$a_\perp=\sqrt{\hbar/m_{\rm p}\omega_{\perp}}$ which gives $\beta_{q_z}=e^{-a_\perp^2 q_z^2/4}$. $\phi_0(r)$
is taken a Wannier function in all other directions thus $\beta_{{\bf q}}=\int d{\bf r}|w_0({\bf r})|^2
e^{i{\bf q}{\bf r}}$  with $w_0({\bf r})$ the Wannier function of the lowest Bloch band. The interaction
Hamiltonian is
\begin{eqnarray}
H_{\rm I}&=&\sum_{{\bf q}}\sum_mc_m^{\dag}c_m e^{i {\bf q}{\bf r}_m^0} \beta_{{\bf q}}\tilde{V}_{\rm cp}({\bf q}){\bf
q}{\bf u}_{{\bf q}}\\ &=&\sum_{m,{\bf q},\lambda}M_{{\bf q},\lambda} e^{i {\bf q}{\bf
r}_m^0}c_m^{\dag}c_m(a_{{\bf q},\lambda}+a^{\dag}_{-{\bf q},\lambda})
\end{eqnarray}
where we can use (\ref{displacement}) to identify
\begin{equation}
M_{{\bf q},\lambda}=(\hbar/2N m_{\rm c}\omega_{{\bf q},\lambda})^{1/2}\beta_{{\bf q}}\tilde{V}_{\rm cp}({\bf q}){\bf
q}{\bf e}_\lambda.\label{theM}
\end{equation}

Extra particles interacting with crystal molecules will force the latter to new equilibrium positions
displaced from their original positions by an extra particle density dependent displacement
\begin{equation}
{\bf v}_j=2\sum_{{\bf q},\lambda}\sqrt{\frac{\hbar}{2N m_{\rm c}\omega_{{\bf q},\lambda}}}\frac{M_{{\bf
q},\lambda}}{\hbar\omega_{{\bf q},\lambda}}\sum_ke^{i {\bf q}({\bf R}_j^0-{\bf
r}_k^0)}c_k^{\dag}c_k.\label{shift2}
\end{equation}
The impact of the displacement is neglected if the relative shift between two neighbouring molecules $|{\bf v}_i-{\bf v}_{i+\delta}|=\langle {\bf v}_i-{\bf v}_{i+\delta}\rangle$  is small compared to the lattice constant.

\section{Correlation functions, corrections and diagonalization of the master equation}\label{AppC}
\subsection{Expectation values and correlation functions}\label{correlations}
The crystal is in a thermal equilibrium at temperature $T$ with a reference state given by
\[\rho_{\rm B}^0=\prod_{{\bf q},\lambda}\exp\Big[\frac{\hbar\omega_{{\bf q},\lambda}}{k_{\rm B}T}a_{{\bf
q},\lambda}^{\dag}a_{{\bf q},\lambda}\Big]\Big(1-\exp\Big[\frac{\hbar\omega_{{\bf q},\lambda}}{k_{\rm
B}T}\Big]\Big).\]
With the displacement of crystal molecules by an extra particle located at site ${\bf r}_j^0$,
\begin{equation}
X_j=\exp[-\sum_{{\bf q},\lambda} u_{{\bf q},\lambda} e^{i{\bf q}{\bf r}_j^0}(a_{-{\bf
q},\lambda}^{\dag}-a_{{\bf q},\lambda})],
\end{equation}
we find
\begin{eqnarray}
 X_{k}^{\dag}X_l &=\exp\Big[\sum_{{\bf q},\lambda}u_{{\bf q},\lambda}\big(e^{i{\bf q}{\bf r}_k^{0}}-e^{i{\bf
 q}{\bf r}_l^{0}}\big)(a_{{\bf q},\lambda}-a_{-{\bf q},\lambda}^{\dag})\Big]\nonumber\\
&=\exp\Big[\sum_{{\bf q},\lambda}\big(Z_{kl}^{{\bf q},\lambda}a_{{\bf q},\lambda}^{\dag}-Z_{kl}^{{\bf
q},\lambda *} a_{{\bf q},\lambda}\big)\Big]=\prod_{{\bf q},\lambda}D(a_{{\bf q},\lambda},Z_{kl}^{{\bf
q},\lambda}),
\end{eqnarray}
where we have introduced $Z_{kl}^{{\bf q},\lambda}=u_{{\bf q},\lambda}\big(e^{-i{\bf q}{\bf
r}_l^{0}}-e^{-i{\bf q}{\bf r}_k^{0}}\big)$ and the displacement operator $D(a,\alpha)\equiv e^{\alpha
a^{\dag}-\alpha^*a}$. In Section~\ref{thermalcrystal} we have introduced the (thermal) expectation value of a
bath operator $O$ by $\langle O\rangle \equiv \tr_{\rm B}\{O\rho_{\rm B}^0\}$ where $\tr_{\rm B}$ denotes the
trace over the bath degrees of freedom. The expectation value of the displacement operator is given by
$\langle D(a_{{\bf q},\lambda},\alpha) \rangle=\exp[-(\bar{n}_{{\bf q},\lambda}(T)+\frac{1}{2})|\alpha|^2]$
where $\bar{n}_{{\bf q},\lambda}(T)=1/(\exp[\hbar\omega_{{\bf q},\lambda}/k_{\rm B}T]-1)$ denotes the thermal
expectation value of the phonon number operator. Therefore we find
\begin{equation}
\langle X_{k}^{\dag}X_l\rangle =e^{-S_T}
\end{equation}
with
\begin{equation}\label{mkoijn}
S_T=2\sum_{{\bf q},\lambda}u_{{\bf q},\lambda}^2\sin\Big[\frac{{\bf q}}{2}({\bf r}_k^{0}-{\bf
r}_l^{0})\Big]^2\big(2\bar{n}_{{\bf q},\lambda}(T)+1\big).
\end{equation}
Notice that the link ${\bf r}_i^0-{\bf r}_j^0$ is always a symmetry axis of the integration interval, the
first Brillouin zone. Since $u_{{\bf q},\lambda}$ and $\bar{n}_{{\bf q},\lambda}(T)$ are invariant under a
rotation with the symmetry of the Brillouin zone the sum over ${\bf q}$ in equation~\eref{mkoijn} becomes
independent of the orientation of that link.

Expectation values of time dependent displacements can we worked out in a similar fashion. With the explicit
form of the bath Hamiltonian $H_{\rm B}=\sum_{{\bf q},\lambda}\hbar\omega_{{\bf q},\lambda}a_{{\bf
q},\lambda}^{\dag}a_{{\bf q},\lambda}$ we can write a bath operator in the interaction picture as
$\tilde{a}_{{\bf q},\lambda}(t)=a_{{\bf q},\lambda}e^{i\omega_{{\bf q},\lambda}t}$ and we find
\begin{eqnarray}
\tilde{X}_k^{\dag}(t)\tilde{X}_l(t)&=&\exp\Big[\sum_{{\bf q},\lambda}\big(Z_{kl}^{{\bf
q},\lambda}\tilde{a}_{{\bf q},\lambda}^{\dag}(t)-Z_{kl}^{{\bf q},\lambda *}\tilde{a}_{{\bf
q},\lambda}(t)\big)\Big]\nonumber\\ &=&\prod_{{\bf q},\lambda}D(a_{{\bf q},\lambda},Z_{kl}^{{\bf q},\lambda}
e^{-i\omega_{{\bf q},\lambda}t}).
\end{eqnarray}
In the following we introduce $\bar{Z}_{ij}^{{\bf q},\lambda}(t)=Z_{ij}^{{\bf q},\lambda}e^{i\omega_{{\bf
q},\lambda}t}$ and use the identity for displacement operators $\prod_{{\bf q},{\bf p}}D(a_{{\bf
q}},\alpha)D(a_{{\bf p}},\beta)=\prod_{{\bf q}}D(\beta,\alpha/2)D(a_{{\bf q}},\alpha+\beta)$ to calculate
\begin{eqnarray}
\fl\langle\tilde{X}_i(t)&\tilde{X}_j(t)\tilde{X}_k(t-\tau)\tilde{X}_l(t-\tau) \rangle=\prod_{{\bf
q},\lambda}\prod_{{\bf p},\kappa}\big\langle D(a_{{\bf q},\lambda},\bar{Z}_{ij}^{{\bf
q},\lambda}(t))D(a_{{\bf p},\kappa},\bar{Z}_{kl}^{{\bf p}\kappa}(t-\tau))\big\rangle\nonumber\\
\fl&=\prod_{{\bf q},\lambda}\langle D(\bar{Z}_{kl}^{{\bf q},\lambda}(t-\tau),\bar{Z}_{ij}^{{\bf
q},\lambda}(t)/2)D(a_{{\bf q},\lambda},\bar{Z}_{ij}^{{\bf q},\lambda}(t)+\bar{Z}_{kl}^{{\bf
q},\lambda}(t-\tau))\rangle\nonumber\\ \fl&=e^{-2S_T}\exp\Big\{-\sum_{{\bf q},\lambda}\Big[\big(\bar{n}_{{\bf
q},\lambda}(T)+1\big)\bar{Z}_{ij}^{{\bf q},\lambda*}(t)\bar{Z}_{kl}^{{\bf q},\lambda}(t-\tau)\nonumber\\
\fl&\hspace{6cm}+\bar{n}_{{\bf q},\lambda}(T)\bar{Z}_{ij}^{{\bf q},\lambda}(t)\bar{Z}_{kl}^{{\bf
q},\lambda*}(t-\tau)\Big]\Big\}\nonumber\\ \fl&=e^{-2S_T}e^{-\Phi_{ij}^{kl}(\tau,T)}.
\end{eqnarray}
Here $\Phi_{ij}^{kl}(\tau,T)$ is found by inserting for $\bar{Z}(t)$ as
\begin{equation}
\Phi_{ij}^{kl}(\tau,T)=\sum_{{\bf q},\lambda}u_{{\bf q},\lambda}^2\big[\big(\bar{n}_{{\bf q},\lambda}+1\big)
g_{ij}^{kl}e^{-i\omega_{{\bf q},\lambda}\tau}+\bar{n}_{{\bf q},\lambda}g_{ij}^{kl}{}^*e^{i\omega_{{\bf
q},\lambda}\tau}\big]\label{bluh}
\end{equation}
with $g_{ij}^{kl}=(e^{-i{\bf q}{\bf r}_j^0}-e^{-i{\bf q}{\bf r}_i^0})(e^{i{\bf q}{\bf r}_l^0}-e^{i{\bf q}{\bf
r}_k^0})$. The quantity $\Phi_{ij}^{kl}(\tau,T)$ depends only on the relative time $\tau = t-t'$ and the two
links ${\bf r}_i^0-{\bf r}_j^0$ and ${\bf r}_k^0-{\bf r}_l^0$. A symmetry argument that relies on the fact
that $({\bf q}{\bf r}^0_i)$ is a projection on a symmetry axis of the integration interval, the first
Brillouin zone, while everything else is an even function in ${\bf q}$ allows us to write (\ref{bluh}) as
\begin{equation}
\Phi_{ij}^{kl}(\tau,T)=\sum_{{\bf q},\lambda}u_{{\bf
q},\lambda}^2\bar{g}_{ij}^{kl}\big[\coth\Big(\frac{\hbar\omega_{{\bf q},\lambda}}{2k_{\rm
B}T}\Big)\cos(\omega_{{\bf q},\lambda}\tau)-i\sin(\omega_{{\bf q},\lambda}\tau)\big]\label{thephi}
\end{equation}
\begin{eqnarray}
\mathwith\bar{g}_{ij}^{kl}=&\cos[{\bf q}({\bf r}_i^0-{\bf r}_k^0)]-\cos[{\bf q}({\bf r}_j^0-{\bf
r}_k^0)]\nonumber\\ &-\cos[{\bf q}({\bf r}_i^0-{\bf r}_l^0)]+\cos[{\bf q}({\bf r}_j^0-{\bf r}_l^0)].
\end{eqnarray}
From this form one can immediately read off the important relations
$\Phi_{ij}^{kl}(\tau,T)=\Phi_{kl}^{ij}(\tau,T)$, $\Phi_{ij}^{kl}(-\tau,T)=\Phi_{ij}^{kl}{}^*(\tau,T)$ and
$|\Phi_{ij}^{kl}(\tau,T)|\leq 2S_T$. Equation~\eref{thephi} is in accordance with the literature
\cite{Mahan}.

For many sites we take the continuum limit and replace the summation over ${\bf q}$ by an integration over
the first Brillouin zone of volume $V_{\rm BZ}$. A variable transformation where $q_x$ is replaced by
$f(q^{d-1},w)$, a function of the remaining momenta $q^{d-1}$ and $w\equiv\omega_{{\bf q},\lambda}$, allows
us to introduce the spectral density
\[J_{ij}^{kl}(w)=V_{\rm BZ}\sum_\lambda\int d{\bf q}^{d-1} \left[\frac{\partial \omega_{{\bf
q},\lambda}}{\partial q_x}\right]^{-1} u_{{\bf q},\lambda}^2\bar{g}_{ij}^{kl}({\bf
q})\Big|_{q_x(q^{d-1},w)},\]
where the $q^{d-1}$ integration interval depends strongly on $w$ and the form of the Brillouin zone while $w$
ranges over all frequencies. In terms of the spectral density equation~\eref{thephi} reads
\begin{equation}
\Phi_{ij}^{kl}(\tau,T)=\int dwJ_{ij}^{kl}(w) \Big[\coth\left( \frac{\hbar w}{2k_{\rm B}T}\right)\cos (w\tau)
-i\sin (w\tau)\Big].
\end{equation}
The bath correlation functions $\xi_{ij}^{kl}(\tau,T)$ that enter the master equation are thus given by
\begin{eqnarray}
\xi_{ij}^{kl}(\tau,T) &= \langle
\tilde{X}_i^{\dag}(t)\tilde{X}_j(t)\tilde{X}_k^{\dag}(t')\tilde{X}_l(t')\rangle - e^{-2S_T}\\
&=e^{-2S_T}(e^{-\Phi_{ij}^{kl}(\tau,T)}-1)\label{xibase}
\end{eqnarray}
and satisfy $\xi_{kl}^{ij}(-\tau,T)=\xi_{ij}^{kl}{}^*(\tau,T)$.
\begin{figure}[t!]
\begin{flushright}
\includegraphics[width=.85\columnwidth]{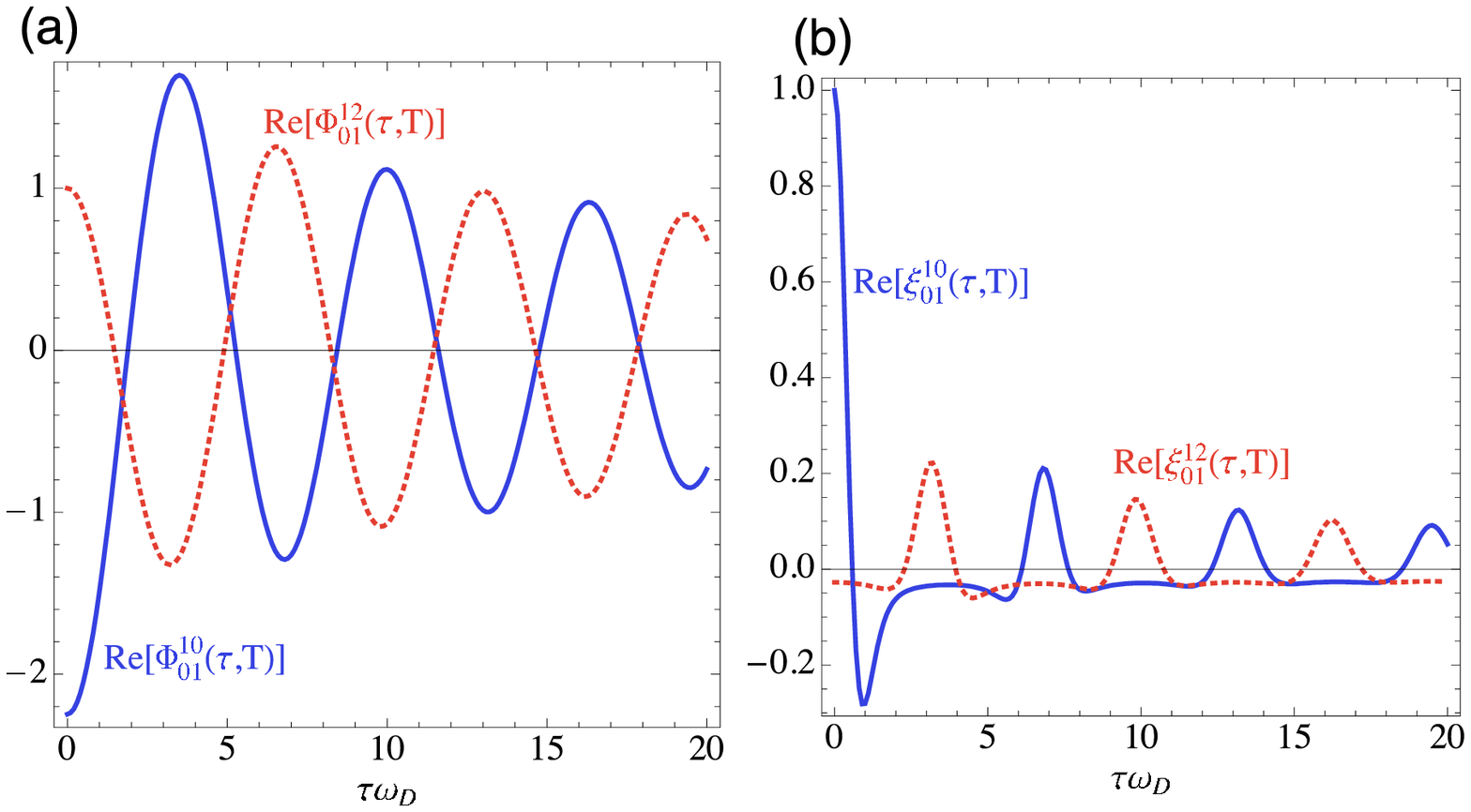}
\end{flushright}
\caption{\label{figXiPhi} Real part of (a) the function $\Phi_{ij}^{kl}(\tau, T)$ and of (b) the correlation function $\xi_{ij}^{kl}(\tau,T)$ for a 1D model with $M_q \propto q/\sqrt{\omega_q}$ and $\omega_q = \omega_{\rm D}\sin(q/2)$ as a function of the (dimensionless) time $\tau\omega_{\rm D}$. Shown are the respective functions for the swapping of a particle, $i = l$ and $k = j$ (solid lines), and the hopping of particles into the same direction, $j=k$ (dashed lines). Notice the overall minimum of the functions $|\Phi_{ij}^{kl}(\tau,T)|$ is attained at  $\tau=0$ for the swap process, i.e. $i = l$ and $k = j$. In panel (b) we chose $S_T=3/4$, while for larger values of $S_T$ the tail of ${\rm Re}[\xi_{01}^{10}(\tau,T)]$, together with all other processes, is strongly suppressed. In the strong coupling limit ($S_T\gg1$) the latter become of the order of $O(e^{-2S_T})$.}
\end{figure}

\subsection{Corrections to the Master Equation}\label{anaResults}
The corrections to the master equation are given by (c.f.~Section~\ref{thermalcrystal})
\begin{eqnarray}
\Gamma_{ij}^{kl}(T)&=&\frac{J^2}{\hbar^2}\int_0^\infty d\tau \Re[\xi_{ij}^{kl}(\tau,T)],\label{aGamma}\\
\Delta_{ij}^{kl}(T)&=&\frac{J^2}{\hbar}\int_0^\infty d\tau \Im[\xi_{ij}^{kl}(\tau,T)],\label{aDelta}\\
\gamma_{ij}^{kl}(T)&=&\frac{J^2}{\hbar^2}\int_0^\infty d\tau \tau\Re[\xi_{ij}^{kl}(\tau,T)].\label{agamma}\\
\delta_{ij}^{kl}(T)&=&\frac{J^2}{\hbar^2}\int_0^\infty d\tau \tau\Im[\xi_{ij}^{kl}(\tau,T)],\label{adelta}
\end{eqnarray}
We can find explicit approximate results to these corrections in the two limiting cases of a weak, $S_T\ll 1$,
and a strong, $S_T\gg 1$, coupling.
\subsubsection{Strong coupling limit}\label{staionaryPhase}
The real part of $\Phi_{ij}^{kl}(\tau,T)$ has a global minimum at $\tau=\tau_0$, see \fref{figXiPhi}(a).
Therefore the function $e^{-\Phi_{ij}^{kl}(\tau,T)}$ is strongly peaked at this minimum, see
\fref{figXiPhi}(b). For a sufficiently strong coupling, $S_T\gg 1$, all contributions to the $\tau$
integration come from a close range around the minimum, $\tau-\tau_0$. Then we may approximate the integral
by performing a stationary phase approximation which features an expansion of $\Phi_{ij}^{kl}(\tau,T)$ around
$\tau_0$. We find
\begin{eqnarray*}
\fl\Re\{\Phi_{ij}^{kl}(\tau,T)\}&=\sum_w J_{ij}^{kl}(w)\coth\Big(\frac{\hbar w}{2k_{\rm
B}T}\Big)\big(\cos(w(\tau-\tau_0))\cos(w\tau)\\ \fl&\hspace{6cm}-\sin(w(\tau-\tau_0))\sin(w\tau_0)\big),\\
\fl&\approx\sum_w J_{ij}^{kl}(w)\coth\Big(\frac{\hbar w}{2k_{\rm
B}T}\Big)\big(\cos(w\tau_0)-w(\tau-\tau_0)\sin(w\tau_0)\\
\fl&\hspace{6cm}-\frac{1}{2}w^2(\tau-\tau_0)^2\cos(w\tau_0)\big),
\end{eqnarray*}
and in the same way
\begin{eqnarray*}
\fl\Im\{\Phi_{ij}^{kl}(\tau,T)\}&\approx\sum_w J_{ij}^{kl}(w)\coth\Big(\frac{\hbar w}{2k_{\rm
B}T}\Big)\big(\sin(w\tau_0)-w(\tau-\tau_0)\cos(w\tau_0)\\
\fl&\hspace{6cm}-\frac{1}{2}w^2(\tau-\tau_0)^2\sin(w\tau_0)\big).
\end{eqnarray*}
This gives for real and imaginary part of the correlation function
\begin{eqnarray}
\fl\begin{array}{c}\Re\\\Im\end{array}[\xi_{ij}^{kl}&(\tau,T)]\approx\exp\Big[\sum_wJ_{ij}^{kl}(w)\coth\Big(\frac{\hbar
w}{2k_{\rm B}T}\Big)\big(\cos(w\tau_0)-1\big)\Big]\nonumber\\
\fl&\times\exp\Big[\sum_wJ_{ij}^{kl}(w)\coth\Big(\frac{\hbar w}{2k_{\rm
B}T}\Big)\Big[w(\tau-\tau_0)\sin(w\tau_0)+\frac{1}{2}w^2(\tau-\tau_0)^2\cos(w\tau_0)\Big]\nonumber\\
\fl&\times\begin{array}{c}\cos\\\sin\end{array}\Big[\sum_wJ_{ij}^{kl}(w)\big(\sin(w\tau_0)+w(\tau-\tau_0)\cos(w\tau_0)\big)\Big]\nonumber
\end{eqnarray}
where real and imaginary part differ only by the trigonometric function at the beginning of the last line in
the last equation. The first line on the right hand side of the last equation does not depend on $\tau$. It exponentially suppresses the rest of the function by
\[
\sum_wJ_{ij}^{kl}(w)\coth\Big(\frac{\hbar w}{2k_{\rm B}T}\Big)\big(\cos(w\tau_0)-1\big).
\]
This factor is of the order of $S_T$ unless $\tau_0 \ll 1$. In the strong coupling limit where $S_T\gg 1$
every process characterized by $J_{ij}^{kl}(w)$ becomes therefore strongly suppressed if the corresponding
correlation is not peaked around $\tau=0$. An analysis of the correlation functions shows that this
condition is only met by correlations to the swap process, $i=l$ and $k=j$. For these processes real and
imaginary part of the correlation functions are given by
\begin{equation}
\fl\begin{array}{c}\Re\\\Im\end{array}[\xi_{ij}^{kl}(\tau,T)]\approx
\exp\Big({\frac{1}{2}\sum_wJ_{ij}^{kl}(w)\coth\left(\frac{\hbar w}{2k_{\rm
B}T}\right)w^2\tau^2}\Big)\times\begin{array}{c}\cos\\\sin\end{array}\Big[\sum_wJ_{ij}^{kl}(w)w\tau\Big]\nonumber
\end{equation}
and the corrections \eref{aGamma}~-~\eref{adelta} are simply Gaussian integrals that calculate as
\begin{eqnarray}
\Gamma_{01}^{10}(T) &\approx&\frac{J^2}{\hbar^2}\frac{\pi^{1/2}}{2}\frac{e^{-B^2/4A_T}}{\sqrt{A_T}},\\
\Delta_{01}^{10}(T) &\approx&\frac{J^2}{\hbar}\frac{\pi^{1/2}}{2}\frac{e^{-B^2/4A_T}}{\sqrt{A_T}}{\rm
Erfi}(B/2\sqrt{A_T}),\\ \gamma_{01}^{10}(T) &\approx&\frac{J^2}{\hbar^2}\left(\frac{1}{2A_T}-\frac{B
e^{-B^2/4A_T} \sqrt{\pi}{\rm Erfi}(B/2\sqrt{A_T})}{4 A_T^{3/2}}\right),\\ \Delta_{01}^{10}(T)
&\approx&\frac{J^2}{\hbar^2}\frac{\pi^{1/2}}{4}\frac{B e^{-B^2/4A_T}}{ A_T^{3/2}},
\end{eqnarray}
where ${\rm Erfi}$ denotes the Error function and the quantities $A_T$ and $B$ are given by
\begin{eqnarray}
A_{T}&\equiv&\int dw\frac{1}{2}J_{01}^{10}(w)w^2\coth\left(\frac{\hbar w}{2k_{\rm B}T}\right),\\
B&\equiv&\int dwJ_{01}^{10}(w)w.
\end{eqnarray}
In the strong coupling limit we can do a simple estimate to show how $\Delta_{01}^{10}(T)$ and
$\gamma_{01}^{10}(T)$ are part of an $J/E_{\rm p}$ expansion for 1D systems. The expansion parameter becomes
visible when including only first order in the series expansion of $-\Phi_{01}^{10}(\tau,T)$ around its
maximum. This is equivalent to the limit of $A_T$ going to zero. In this limit we find
\begin{eqnarray}
\Delta_{01}^{10}(T) &\approx \frac{J^2}{\hbar\omega_{\rm D} B},\\
\gamma_{01}^{10}(T) &\approx \frac{J^2}{(\hbar\omega_{\rm D} B)^2}.
\end{eqnarray}
The quantity $B$ for this process reads
\begin{eqnarray}
B &= \int dw J_{01}^{10}(w) w = \frac{2}{\pi}\int dq \Big(\frac{M_q}{\hbar\omega_q}\Big)^2 \sin(q/2)^3\\
&\approx \frac{1}{\pi\hbar\omega_{\rm D}}\int dq \frac{M_q^2}{\hbar\omega_q} [1-\cos(q)]=\frac{\tilde{V}_{i,i}^{(1)}-\tilde{V}_{i,i+1}^{(1)}}{2\pi\hbar\omega_{\rm D}}.
\end{eqnarray}
In the two models in Section~\ref{rdc} we find that the onsite phonon mediated interaction
$\tilde{V}_{i,i}^{(1)}= 2E_{\rm p}$ dominates over all the others. Therefore we can approximately write
$B\approx E_{\rm p}/\pi\hbar\omega_{\rm D}$ with which we find
\begin{eqnarray}
\Delta_{01}^{10}(T) &\approx \pi\frac{J^2}{E_{\rm p}},\\
\gamma_{01}^{10}(T) &\approx \pi^2\frac{J^2}{E_{\rm p}^2}.
\end{eqnarray}
\subsubsection{Weak coupling limit}
In the weak coupling limit, $S_T \ll 1$, we use the identity $\Phi_{ij}^{kl}(\tau,T)\leq 2S_T$ to perform a
pointwise expansion of the correlation functions \eref{xibase} in $\Phi_{ij}^{kl}(\tau,T)$ as
\begin{equation}
\xi_{ij}^{kl}(\tau,T)\approx e^{-2S_T}\Phi_{ij}^{kl}(\tau,T).
\end{equation}
With this expansion real and imaginary part of $\xi_{ij}^{kl}(\tau,T)$ simply become
\begin{eqnarray}
\Re[\xi_{ij}^{kl}(\tau,T)]&=&e^{-2S_T}\int dwJ_{ij}^{kl}(w)\coth\left( \frac{\hbar w}{2k_{\rm B}T}\right)\cos
(w\tau),\\
\Im[\xi_{ij}^{kl}(\tau,T)]&=&-e^{-2S_T}\int dwJ_{ij}^{kl}(w)\sin (w\tau),
\end{eqnarray}
where the $w$ integration ranges over all frequencies. This way the Debye frequency of our model functions as
a natural cutoff (compare \cite{Breuer}). The functional identities
\begin{eqnarray*}
\int_0^\infty d\tau \cos(w \tau) &=& \pi\delta(w),\\
\int_0^\infty d\tau \sin(w \tau) &=& P(1/w),\\
\int_0^\infty d\tau \tau\cos(w \tau) &=& -\pi\delta'(w),\\
\int_0^\infty d\tau \tau\sin(w \tau) &=& -P(1/w^2),
\end{eqnarray*}
with $P(x)$ the Cauchy principal value of $x$ and the delta functional $\delta(x)$, allow us to find the corrections in the weak coupling limit as
\begin{eqnarray}
\Gamma_{ij}^{kl}(\tau,t) &=&\frac{\tilde{J}^2}{\hbar^2}\pi\Big[J_{ij}^{kl}(w)\coth\Big(\frac{\hbar w}{2k_{\rm
B}T}\Big)\Big]_{w=0},\\
\Delta_{ij}^{kl}(\tau,t) &=&\frac{\tilde{J}^2}{\hbar}\lim_{\epsilon\rightarrow 0}\int
dw J_{ij}^{kl}(w)\frac{w}{w^2+\epsilon^2},\\
\gamma_{ij}^{kl}(\tau,t) &=&-\frac{\tilde{J}^2}{\hbar^2}\lim_{\epsilon\rightarrow 0}\int dw J_{ij}^{kl}(w)\coth\Big(\frac{\hbar
w}{2k_{\rm B}T}\Big)\frac{\epsilon^2-w^2}{(\epsilon^2+w^2)^2}\label{ijnijnijnijn},\\
\delta_{ij}^{kl}(\tau,t)&=&-\frac{\tilde{J}^2}{\hbar^2}\pi\partial_wJ_{ij}^{kl}(w)|_{w=0},
\end{eqnarray}
where we have written out the Cauchy principal value integrals. We remark that the integration
\eref{ijnijnijnijn} becomes infinite in the limit of zero temperature.
\subsection{Diagonalization of the master equation}\label{diagonalization}
As noted above the corrections to the master equation $\Gamma_{ij}^{kl}(T), \Delta_{ij}^{kl}(T),
\gamma_{ij}^{kl}(T)$ and $\delta_{ij}^{kl}(T)$ are just matrix entries and have little physical meaning by
themselves.

In this section we show how to estimate the energy of the corrections to the master equation \eref{mastaend} for
a single particle at many sites. We start by treating the self energies,
\begin{eqnarray}
\sum_{\langle ij\rangle \langle
kl\rangle}\Delta_{ij}^{kl}(T)b_{ij}b_{kl}&=\sum_{i}\sum_{m,n}\Delta_{i,i+m}^{i+m,i+m+n}c_i^\dag c_{i+m+n}\\
&=\sum_{\bf q}\sum_{m,n}\Delta_{0,m}^{m,m+n}e^{i {\bf q}({\bf r}_m^0+{\bf r}_n^0)}c_{\bf q}^\dag c_{\bf
q}\label{c40}
\end{eqnarray}
where the sums over $m,n$ range over basis vectors in the lattice. In the second step we have made use of the
fact that the energies $\Delta_{i,i+m}^{i+m,i+m+n}$ do not depend on the index $i$ and that in ${\bf q}$
space we can write $\sum_i c_i^\dag c_{i+m}=\sum_{\bf q} e^{i {\bf q}{\bf r}_m^0}c_{\bf q}^\dag c_{\bf q}$.
The eigenvalues to \eref{c40} are simply given by
\begin{equation}
\Delta_{\bf q} = \sum_{m,n}\Delta_{0,m}^{m,m+n}e^{i {\bf q}({\bf r}_m^0+{\bf r}_n^0)}
\end{equation}
and the largest eigenvalue gives an upper bound to the energy of this term in the master equation.

To determine the energy of the thermal fluctuations we notice that this term can be written as
\begin{equation}
\frac{1}{\hbar}\sum_{\langle ij\rangle\langle kl\rangle}\hbar\Gamma_{ij}^{kl}(T)\big(
\{b_{ij}b_{kl},\rho_{\rm S}(t)\} - 2b_{kl}\rho_{\rm S}(t)b_{ij}\big)
\end{equation}
where we have introduced $b_{ij}= c_i^\dagger c_j$ and we estimate its amplitude by the energy of
$\sum_{\langle ij\rangle\langle kl\rangle}\hbar\Gamma_{ij}^{kl}(T) b_{ij}b_{kl}$. Since $\Gamma_{ij}^{kl}(T)$
has the same properties as $\Delta_{ij}^{kl}(T)$ the eigenvalues are given by
\begin{equation}
\Gamma_{\bf q}=\sum_{m,n}\Gamma_{0,m}^{m,m+n}e^{i {\bf q}({\bf r}_m^0+{\bf r}_n^0)}.
\end{equation}
The dissipative term proportional to $\gamma_{ij}^{kl}(T)$ in the single particle limit is given by
\begin{eqnarray}
\fl-\frac{i\tilde{J}}{\hbar}\sum_{\langle ij\rangle\langle kl\rangle} \gamma_{ij}^{kl}(T) \Big[
b_{ij}\Big(\sum_{{k'}}b_{k'l}-\sum_{l'}b_{kl'}\Big)\rho_{\rm S}(t) +\rho_{\rm
S}(t)\Big(\sum_{{k'}}b_{k'l}-\sum_{l'}b_{kl'}\Big)b_{ij}\nonumber\\ - b_{ij}\rho_{\rm
S}(t)\Big(\sum_{{k'}}b_{k'l}-\sum_{l'}b_{kl'}\Big) - \Big(\sum_{{k'}}b_{k'l}-\sum_{l'}b_{kl'}\Big)\rho_{\rm
S}(t)b_{ij}\Big],\label{nonumba}
\end{eqnarray}
where $k'$ and $l'$ denote the nearest neighbours of $k$ and $l$. We can write the first two terms as
\begin{eqnarray}
\fl\sum_{\langle ij\rangle\langle kl\rangle} \gamma_{ij}^{kl}(T)
b_{ij}\Big(\sum_{{k'}}b_{k'l}-\sum_{l'}b_{kl'}\Big)=\sum_{i}\sum_{m,n,o}\Big(\gamma_{i,i+m}^{i+m+n,i+m+n+o}(T)\nonumber\\
\hspace{4cm}-\gamma_{i,i+m}^{i+m,i+m+n+o}(T)\Big)c_i^\dag c_{i+m+n+o},\\ \fl\sum_{\langle ij\rangle\langle
kl\rangle} \gamma_{ij}^{kl}(T)
\Big(\sum_{{k'}}b_{k'l}-\sum_{l'}b_{kl'}\Big)b_{ij}=\sum_{k}\sum_{m,n,o}\Big(\gamma_{k+m+n,k+m+n+o}^{k,k+m+n}\nonumber\\
\hspace{4cm}-\gamma_{k+m+n,k+m+n+o}^{k,k+m}\Big)c_k^\dag c_{k+m+n+o},
\end{eqnarray}
and as before the sums over $m,n,o$ range over all basis vectors. The property $\gamma_{ij}^{kl}(T) =
\gamma_{kl}^{ij}(T)$ allows us to collect the first two terms in \eref{nonumba} by an anticommutator
\begin{equation}
\fl-\frac{i\tilde{J}}{\hbar}\Big\{\sum_{m,n,o}\Big(\gamma_{0,m}^{m+n,m+n+o}(T)-\gamma_{0,m}^{m,m+n+o}(T)\Big)\sum_ic_i^\dag
c_{i+m+n+o},\rho_{\rm S}(t)\Big\}.
\end{equation}
In this case the amplitude of the entire term can be estimated by the eigenvalues of the operator inside of
the anti commutator. They are given by
\begin{equation}
\gamma_{\bf q}=\sum_{m,n,o}\Big(\gamma_{0,m}^{m+n,m+n+o}(T)-\gamma_{i,i+m}^{m,m+n+o}(T)\Big)e^{i {\bf q}({\bf
r}_m^0+{\bf r}_n^0+{\bf r}_o^0)}.
\end{equation}
We use the same line of argumentation to estimate the eigenvalues of the dissipative correction proportional
to $\delta_{ij}^{kl}$ and find
\begin{equation}
\delta_{\bf q}=\sum_{m,n,o}\Big(\delta_{0,m}^{m+n,m+n+o}(T)-\delta_{0,m}^{m,m+n+o}(T)\Big)e^{i {\bf q}({\bf
r}_m^0+{\bf r}_n^0+{\bf r}_o^0)}.
\end{equation}

In a one-dimensional system the extra particles locate at sites of postion $r_m^0=m a$ with the lattice
constant $a$. There are only the two basisvectors $\pm a$ and therefore the indices $m,n$ and $o$ range only
over $\pm 1$. This gives
\begin{eqnarray}
\Gamma_q(T)&=\sum_{m,n}\Gamma_{0,m}^{m,m+n}(T)e^{i {\bf q}({\bf r}_m^0+{\bf r}_n^0)}\\
&=\Gamma_{0,1}^{1,2}(T)
e^{i2qa}+\Gamma_{0,1}^{1,0}(T)+\Gamma_{0,-1}^{-1,0}(T)+\Gamma_{0,-1}^{-1,-2}(T)e^{-i2qa}\\
&=2\Gamma_{0,1}^{1,0}(T)+2\Gamma_{0,1}^{1,2}(T)\cos(qa).
\end{eqnarray}
In the second step we have used the fact that the origin of the indices $i,j,k,l$ is the function
$g_{ij}^{kl}(q)$ and therefore the corrections show the same symmetries with respect to the indices as
$g_{ij}^{kl}(q)$ itself. In the same way we find the other eigenvalues as
\begin{eqnarray}
\Delta_q(T) &=  2 [\Delta_{01}^{10}(T)+\Delta_{01}^{12}(T)\cos(qa)],\\ \gamma_q(T) &=
2[(\gamma_{01}^{23}(T)-\gamma_{01}^{13}(T))\cos(3qa)\nonumber\\
&\hspace{.5cm}+(\gamma_{01}^{21}(T)+\gamma_{01}^{01}(T)+\gamma_{01}^{0,-1}(T)-\gamma_{01}^{1,-1}(T))\cos(qa)],\\
\delta_q(T) &= 2[(\delta_{01}^{23}(T)-\delta_{01}^{13}(T))\cos(3qa)\nonumber\\
&\hspace{.5cm}+(\delta_{01}^{21}(T)+\delta_{01}^{01}(T)+\delta_{01}^{0,-1}(T)-\delta_{01}^{1,-1}(T))\cos(qa)].
\end{eqnarray}

\end{document}